\documentclass[superscriptaddress,twocolumn,secnumarabic,amssymb,nobibnotes,aps,prb]{revtex4-2}
\usepackage{siunitx}
\usepackage{graphicx}
\usepackage{amsmath}
\usepackage{amsfonts}
\usepackage{xcolor}

\newcommand{\Tgb}{$T_{\text{g,}\beta} $}

\newcommand{\Tg}{$T_{\text{g}} $}
\newcommand{\w}{\,wt.\%}
\newcommand{\Tstar}{$T^* $}

\begin{document}

\title{Terahertz Dynamics in the Glycerol-Water System}

\author{Johanna K\"{o}lbel}
\affiliation{Department of Chemical Engineering, University of Cambridge, Cambridge, CB3 0AS, UK}

\author{Walter Schirmacher}
\affiliation{Department of Physics, Johannes Gutenberg University, 55099 Mainz, Germany}
\affiliation{Center for Life Nano Science @Sapienza, Istituto Italiano di Tecnologia, 295 Viale Regina Elena, I-00161, Roma, Italy}

\author{Evgenyi Shalaev}
\affiliation{AbbVie, Irvine, CA, USA}

\author{J. Axel Zeitler}
\email{jaz22@cam.ac.uk}
\affiliation{Department of Chemical Engineering, University of Cambridge, Cambridge, CB3 0AS, UK}

\date{\today}

\begin{abstract}
The model glass-former glycerol and its aqueous mixtures were investigated with terahertz-time domain spectroscopy (THz-TDS) in the frequency range of \SIrange{0.3}{3.0}{THz} at temperatures from \SIrange{80}{305}{K}. It was shown that the infrared absorption coefficient measured with THz-TDS can be theoretically related to the reduced Raman intensity ($\propto \alpha/\omega^2$) and the reduced density of states ($\propto \alpha/\omega^3$) and the agreement with experimental results confirms this. The data were further used to investigate the behaviour of model glasses in the harmonic (below the glass transition temperature \Tg), anharmonic (above \Tg), and liquid regime. The onset temperature of the molecular mobility as measured by the infrared active dipoles, \Tg, was found to correlate with the onset of anharmonic effects, leading to an apparent shift of the boson peak and obscuring it at elevated temperatures. The influence of clustered and unclustered water on the dynamics, the boson peak, and the vibrational dynamics was also investigated.  A change in structural dynamics was observed at a water concentration of approximately \SI{5}{\w}, corresponding to a transition from isolated water molecules distributed homogeneously throughout the sample to the presence of small water clusters and an increased number of water-water hydrogen bonds which lower the barriers on the potential energy surface. 
\end{abstract}
\maketitle

\section{Introduction}
\subsection{Dynamics of Glass-Formers}

Upon cooling an organic molecular liquid to temperatures below its melting point, it  either forms a crystal or remains disordered, ultimately forming an amorphous solid. Initially, at temperatures just below its freezing temperature a liquid is referred to as supercooled. In such thermodynamically metastable systems the Stokes-Einstein relation breaks down, and diffusion and viscosity decouple \cite{corsaro2019stokes}. The molecular arrangement remains similar to that of liquids and long-range order is completely absent. Upon further cooling to temperatures below the glass transition temperature (\Tg), a sharp change in the thermal expansion coefficient and specific heat is observed and a glass is formed. 

Motions that occur at temperatures at and above \Tg\ include translational, cooperative molecular motions. At temperatures below \Tg\ the molecular motions are restricted to local or so-called secondary mobility of atoms rattling in the cages formed by their neighbours \cite{gotze1998essentials}. Experimentally, the \Tg\ can be measured by a range of techniques, such as differential scanning calorimetry (DSC), which  detects the change in heat capacity at the glass transition temperature. 

Terahertz time-domain spectroscopy (THz-TDS) probes frequencies in the range of \SIrange{0.35}{3}{\tera\hertz} by coupling to dipole moments at photon energies on the order of the hydrogen bond strength and picosecond relaxation time. It hence provides a sensitive probe of mobility in condensed phase organic molecular materials. Past studies using THz-TDS have investigated the complex interplay between reorientation motion and vibration dynamics in hydrogen-bonded liquids \cite{yomogida2010comparative, yomogida2010comparative2, yomogida2010comparative3}.

Due to the inherent disorder of glasses, no discrete phonon modes can be sustained and the coupling of photons to the vibrational density of states (VDOS) results in the relatively featureless so-called microscopical peak that constitutes the main absorption mechanism in the terahertz range \cite{taraskin2006universal} with a frequency-squared dependence as predicted by Debye theory \cite{sibik2016direct}. With increasing temperature, an overall increase in absorption is measured within the accessible spectral bandwidth of THz-TDS instruments ($< \SI{3}{THz}$).  In addition to the microscopical peak, THz-TDS also detects the excess density of states (DOS) above the Debye level. This excess DOS is termed boson peak (BP).

While the exact origin of the BP is still not fully understood, it is generally accepted that the BP is a harmonic phenomenon due to inherent disorder, and should therefore be temperature independent \cite{schirmacher1998harmonic, marruzzo2013vibrational, schirmacher2015theory}.

Goldstein proposed that properties of glasses can be explained by the characteristics of the potential energy surface (PES) \cite{goldstein1969viscous}. The atomic dynamics have been described as a consequence of the shape of the PES \cite{ruggiero2017significance}. In this context, at low temperatures, the system is trapped in a deep and steep minimum which can be approximated as harmonic. Anharmonic effects play only a small role at the low temperatures. The measurable increase in absorption coefficient at single frequencies, however, is due to anharmonic effects \cite{markelz2007protein}. At temperatures at which the apparent centre frequency of the BP is not affected by anharmonicity, we will call the temperature range the harmonic regime.

Once the temperature is increased sufficiently for the system to escape the deep minimum, it has access to shallower regions of the PES that are characterised by numerous local minima separated by smaller energy barriers. The temperature at which the system contains sufficient thermal energy to leave the deep minimum on the PES coincides with \Tstar, i.e. the onset of local mobility, observed by THz-TDS as an increase in the rate of absorption change with temperature. The beta-relaxation is the result of hopping processes across low energy barriers. Anharmonic effects begin to dominate the spectra above a certain temperature when non-harmonic parts of the PES minima can be explored, resulting in an apparent change of the BP frequency.

A further temperature increase, ultimately resulting in \Tg, opens up more local minima on the PES while anharmonic effects continue to play an important role. The larger the amplitude of the possible motions, the higher the associated change in dipole moment and hence the measured absorption coefficient. The shallower the minima, the more anharmonicity influences the spectra.

 Glycerol is a widely studied network material and model system for a glass former \cite{wuttke1994neutron} as it is non toxic and easily supercools and has been investigated with experimental techniques like Raman scattering \cite{wang1971raman1, wang1971raman2}, infrared spectroscopy \cite{dashnau2006hydrogen, perova1998far}, dielectric spectroscopy \cite{murthy2000experimental, huck1988dielectric}, neutron and light scattering \cite{wuttke1994neutron, dirama2005role}, DSC \cite{angell1982test}, terahertz spectroscopy \cite{sibik2013terahertz}, as well as theoretical work with molecular dynamics simulations \cite{dashnau2006hydrogen, busselez2009molecular}. Its \Tg\ was found to lie between \SI{185}{\kelvin} and \SI{194}{\kelvin} \cite{angell1982test, schneider1998dielectric, ryabov2003features, sibik2014thermal}.
In glycerol, the VDOS extends from approximately \SIrange{0.5}{7}{\tera\hertz} \cite{schneider1998dielectric}.

THz-TDS measurements of glycerol revealed that its \Tgb\ is more pronounced than the \Tg \cite{sibik2014thermal, capaccioli2015coupling}. Glycerol is hence a promising model system to investigate whether anharmonic effects can be detected at temperatures even below \Tg. Here we study the impact of temperature changes on glycerol dynamics, especially anharmonic effects that can obscure the BP.

The density of states of pure glycerol, including the elevated temperature range,
where the spectrum becomes temperature dependent, is well documented: \citet{chumakov2004collective} embedded a probe molecule in a glass matrix and monitored its translational motions with nuclear inelastic scattering. Because the probe followed the collective motions of the glass with a correlation length larger than the probe size, the density of states of collective motions of the glass matrix, i.e. the VDOS multiplied with the Debye-Waller factor, was directly measured. It was reported that a significant part of the BP in the model glass former glycerol is constituted of collective modes and that it disappears close to \Tg\ due to increased sample mobility. They further observed an exponential decrease in the reduced VDOS at energies above the BP maximum \cite{chumakov2004collective}.

Wuttke \textit{et al.} found that the reduced density of states of glycerol is temperature-independent at frequencies above the BP maximum \cite{wuttke1995fast}. At frequencies below the maximum, a temperature-dependent change was observed even at temperatures below \Tg. These temperature-dependent changes in the BP occur above the characteristic temperature \Tstar, which coincides with the onset of localised motions and therefore increased anharmonicity.  The onset of localised motions at \Tstar\ is typically too subtle to be detected in DSC measurements \cite{sibik2015predicting}. Ruggiero et al. described \Tstar\ as the  vitrification of the $\beta$-relaxation process as originally detailed by Johari and Goldstein, also called \Tgb. \cite{ruggiero2017significance}
\subsection{Glycerol-Water Mixtures}

Modern biopharmaceutical drugs are often developed into complex amorphous formulations that are prepared by spray-drying or freeze-drying (lyophilization). Given its resistance to crystallisation, glycerol is used widely as a so-called cryoprotectant of cells and organs and also to stabilise lyophilised proteins \cite{sztein2018history, cicerone2015stabilization}. Some pharmaceutical degradation mechanisms are catalysed by water clusters and the influence of water content on the stability of lyophilised products has been investigated previously \cite{shalaev1996does}. In macromolecules, water can both serve to stabilise the native structure while simultaneously acting as a catalyst for destabilisation. Starciuc et al. suggested that unclustered water molecules (that are commonly found in products containing only a few wt.\,\% water) are less catalytically active than water clusters \cite{starciuc2021water}. They hypothesised that proton transfer becomes possible only once a water clustering threshold is exceeded, thereby supporting pharmaceutical degradation reactions such as amide hydrolysis and deamidation.

The model system of glycerol-water mixtures has been studied widely \cite{murata2012liquid, dashnau2006hydrogen, murthy2000experimental, towey2012structural, towey2011structure, towey2012molecular}:
Towey \textit{et al.} used a combination of neutron diffraction experiments and computational modelling to investigate the structure and hydrogen-bonding of pure glycerol \cite{towey2011structure}, dilute aqueous glycerol solution \cite{towey2012structural}, and glycerol-water mixtures for glycerol mole fractions of  $x_\text{g} = 0.05, 0.10, 0.25, 0.50, 0.80, \text{ and } 1.00$ \cite{towey2012molecular}. They found bipercolating clusters of both water and glycerol in samples containing a mole fraction between 0.75 and 0.5 water (i.e. between \SIrange{16.4}{37}{\w} water). They further found that water maintained its full hydrogen-bonding capacity independently of the glycerol concentration.

Murata and Tanaka found evidence for a water liquid-liquid transition (LLT) without macroscopic phase separation in glycerol-water mixtures containing between approximately \SIrange{45.5}{56}{\w} water \cite{murata2012liquid}.

Recently, Starciuc et al. investigated glycerol-water mixtures containing between \SIrange{0}{40}{\w} water with low- and high-wavenumber Raman spectroscopy, and three water-content regions were found \cite{starciuc2021water}. This confirmed the existence of a threshold for water clustering in glycerol.  A transition from predominantly unclustered water molecules to small water clusters was reported at $6.1 \pm 0.7$\,\w water and a second threshold at $18.6 \pm 4.4$\,\w water, which was proposed to correspond to the formation of large water clusters associated with the onset of freezing. 

And while glycerol and glycerol-water systems have been investigated thoroughly with a range of methods, there has not yet been a comprehensive study in the terahertz frequency range over a range of temperatures to investigate the role of water, especially the formation of water clusters, in glycerol-water mixtures with a water content below \SI{30}{\w water}, where past studies have shown the existence of several water regions. 
Typical specification limits for freeze-dried pharmaceuticals are below \SI{5}{\w}. By understanding the structural dynamics of the glycerol-water system, such water content specifications in the design of pharmaceutical products can be scientifically substantiated and their design can be optimised. 

The dynamics of the model system of glycerol-water mixtures are investigated with THz-TDS by varying both the temperature and the water concentration. The aim was to extract as much information as possible about both the glass and the amorphous solid and to investigate the role of the BP and the shape of the VDOS and their relationship with the glass transition temperatures.

In addition to the THz-TDS data at various (cryogenic) temperatures, high-quality room temperature data of liquid glycerol-water mixtures with different water concentrations were also acquired to further understand how the water content influences the terahertz spectra of the liquids.

\section{Materials and Methods}
\subsection{Variable-Temperature THz-TDS Measurements}

Glycerol was obtained from Sigma Aldrich (Poole, UK) and was mixed with Milli-Q water (IQ 7000, Merck, Darmstadt, Germany, resistivity \SI{18.2}{\mega\ohm\centi\meter}) in various ratios (\SIrange{0}{30}{\w water}). After mixing, the samples were sealed and left standing or briefly de-gassed inside a desiccator attached to a vacuum pump. No experimental differences between samples prepared by the two methods were observed. For subsequent measurements, a drop of liquid sample was sealed between two z-cut quartz windows (Crystran, UK) separated by a \SI{100}{\micro\meter} thick PTFE spacer in a liquid cell which was then fit into the cryostat and placed under vacuum. The samples were analysed in transmission using a commercial Terapulse 4000 spectrometer (TeraView Ltd., Cambridge, UK) with a liquid nitrogen cryostat attached, as described by Sibik et al. \cite{sibik2013glassy}. Two z-cut quartz windows (Crystran, UK) were used for reference measurements. The spectral range the instrument was able to access was in the range of \SIrange{0.35}{3}{\tera\hertz}, depending on sample absorption.

Utilising the cryostat and liquid nitrogen cooling, a wide range of temperatures were studied (\SIrange{80}{305}{\kelvin}). At each temperature, a co-average of 1000 waveforms was acquired for both reference and sample and used to calculate the optical constants.

To increase reproducibility and facilitate comparison between different water concentrations, the measurements were controlled with a computer program that kept the length of temperature steps and therefore the overall heating rate constant at about \SI{1}{\kelvin\per\minute}.

\subsection{THz-TDS Measurements at Room Temperature}

Glycerol and its mixtures were prepared following the same steps as described above. However, instead of sealing the sample into a liquid cell that fit into the cryostat, a room temperature liquid cell was utilised which was easier to assemble. This liquid cell also consisted of two z-cut quartz windows separated by a \SI{100}{\micro\meter} thick PTFE spacer. This was subsequently placed into the measurement chamber which was purged with dry nitrogen and samples were analysed in transmission using the Terapulse 4000 spectrometer utilising the same settings as for measurements at variable temperatures.

\subsection{Relationship between reduced density of states measured with THz-TDS, neutron,  Raman, and nuclear inelastic scattering}

In this section we develop theoretical ideas relating the infrared absorption coefficient $\alpha(\omega)$ to the mechanical degrees of freedom. The absorption coefficient is related to the dielectric loss function by $\alpha(\omega) = \omega \epsilon''(\omega) / \left( n(\omega) c \right) $. The loss function, in turn, can be related to the complex shear modulus $G(\omega) = G'(\omega) - i G''(\omega)$ by the formula of Gemant \cite{dimarzio1974connection, gemant1935conception,jakobsen2005dielectric}, as utilised by Schirmacher, Ruocco and Mazzone \cite{schirmacher2016theory}.

\begin{equation}
\epsilon''= \mathfrak{Im} \bigg \{  \frac{\textstyle 1}{\textstyle 1+\frac{VG}{kT}}  \bigg \}  \propto \mathfrak{Im} \bigg \{  \frac{\textstyle 1}{\textstyle G } \bigg \}
\end{equation}

$G(\omega)$ is the complex mechanical shear modulus, the frequency dependence of which is mainly given by the sound attenuation of the shear waves $\Gamma(\omega)$  \cite{marruzzo2013heterogeneous}:

\begin{equation}
G(\omega) = G'(\omega) \left[ 1- i \Gamma(\omega)/ \omega \right]
\end{equation}

Therefore $\alpha(\omega)$ is given by

\begin{equation}
\alpha(\omega) \propto  \omega \epsilon''(\omega) \propto \omega^2\frac{\Gamma(\omega)}{1+\Gamma(\omega)^2/\omega^2}
\end{equation}

This means that we can extract the transverse sound attenuation coefficient from the infrared absorption, which is usually not accessible to experimental investigation. 

A very similar formula can be obtained for the Raman intensity $I_{\text{AB}}(\omega) = \left[ n(\omega)+1 \right] \chi''_{\text{AB}}(\omega)$, where $\text{A}$ and $\text{B}$ denote $\text{V}$ or $\text{H}$ polarisation and $n(\omega)=(\exp{(\hbar \omega/(\text{k}T))}-1)^{-1}$. The Raman susceptibilities can be written as \cite{schirmacher2015theory}:

\begin{equation}
\chi''_{\text{VV}}(\omega) = \frac{4}{3} \chi''_{\text{VH}}(\omega) + Af_1 \chi^\xi_{\text{L}}(\omega)
\end{equation}

\begin{equation}
\chi''_{\text{VH}}(\omega) = A f_2 \frac{1}{30} \left( 2\chi^\xi_{\text{L}}(\omega) + 3 \chi^\xi_{\text{T}}(\omega) \right)
\end{equation}

where 

\begin{eqnarray}
\chi^\xi_{\text{L,T}}(\omega) &= &\mathfrak{Im} \bigg \{ \frac{3}{k^3_\xi} \int_0^{k_\xi} k^2 dk \frac{\textstyle k^2}{\textstyle -\omega^2 + k^2 c^2_{\text{L,T}}(\omega)}  \bigg \} \nonumber\\
&=&
\mathfrak{Im} \bigg \{
\frac{1}{c^2_{\text{L,T}}(\omega)} + O(\omega^2)
\bigg \}
\end{eqnarray}

These equations have been derived under the assumption of fluctuating Pockels constants \cite{martin1974model, schmid2008raman}. The upper cut-off $k_\xi$ is proportional to the inverse correlation length $\xi$ of the Pockels constant fluctuations. $c_{\text{L}}(\omega)$ and $c_{\text{T}}(\omega)$ are the complex longitudinal and transverse sound velocities. $f_1$ and $f_2$ are the pre-factors of the longitudinal and transverse Pockels correlation functions, respectively. In terms of the longitudinal modulus $M(\omega)$ and the shear modulus $G(\omega)$ the complex sound velocities are given by

\begin{equation}
\rho c^2_{\text{L}}(\omega) = M(\omega) \quad \quad  
\rho c^2_{\text{T}}(\omega) = G(\omega) 
\end{equation}

where $\rho$ is the mass density.

Because the inverse of the transverse sound velocity is much larger than the longitudinal one, the Raman scattering intensity will be dominated by the transverse susceptibility. For low enough frequencies we essentially have:
\begin{equation}
I_{\text{AB}}(\omega) \propto \left[ n(\omega)+1 \right] \mathfrak{Im} \bigg \{ \frac{\textstyle 1}{\textstyle G } \bigg \}.
\end{equation}

So we expect the reduced Raman intensity to be proportional to the infrared absorption coefficient divided by $\omega^2$:

\begin{equation}\label{equ:raman1}
\tilde{I}(\omega) \equiv \frac{ I(\omega) }{\omega( n(\omega) + 1)} \propto  \alpha(\omega) / \omega^2.
\end{equation}

The reduced Raman intensity is also commonly represented as: \cite{schmid2008raman} 
\begin{equation}\label{equ:reducednew}
\tilde{I}(\omega) \propto C(\omega)\frac{g(\omega)}{\omega^2}
\end{equation}
with the phenomenological frequency dependent coupling coefficient $C(\omega)$ and the reduced DOS $\frac{g(\omega)}{\omega^2}$. In most materials $C(\omega)\propto\omega$ \cite{surovtsev2002frequency}.

Thus, the reduced DOS $\frac{g(\omega)}{\omega^2}$ is related to the infrared absorption coefficient by
\begin{equation}\label{equ:reducednew2}
\frac{g(\omega)}{\omega^2} \propto \frac{\alpha(\omega)}{\omega^3}
\end{equation}

\section{Results and Discussion}
\subsection{Pure Glycerol Measurements and Analysis}

\subsubsection{The Boson Peak}
\begin{figure*}
\centering
  	\includegraphics[width=0.9\columnwidth]{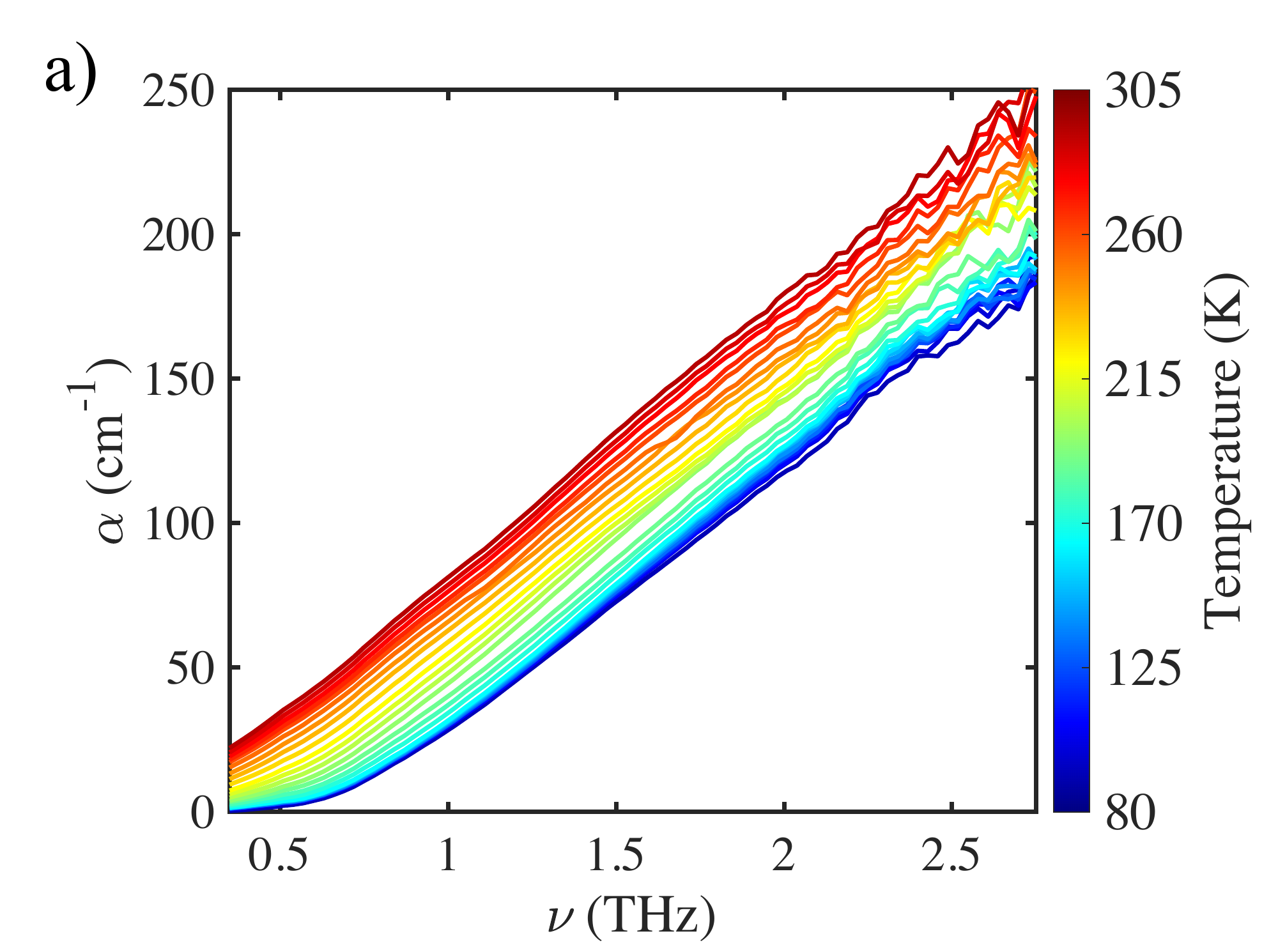}
  	\includegraphics[width=0.9\columnwidth]{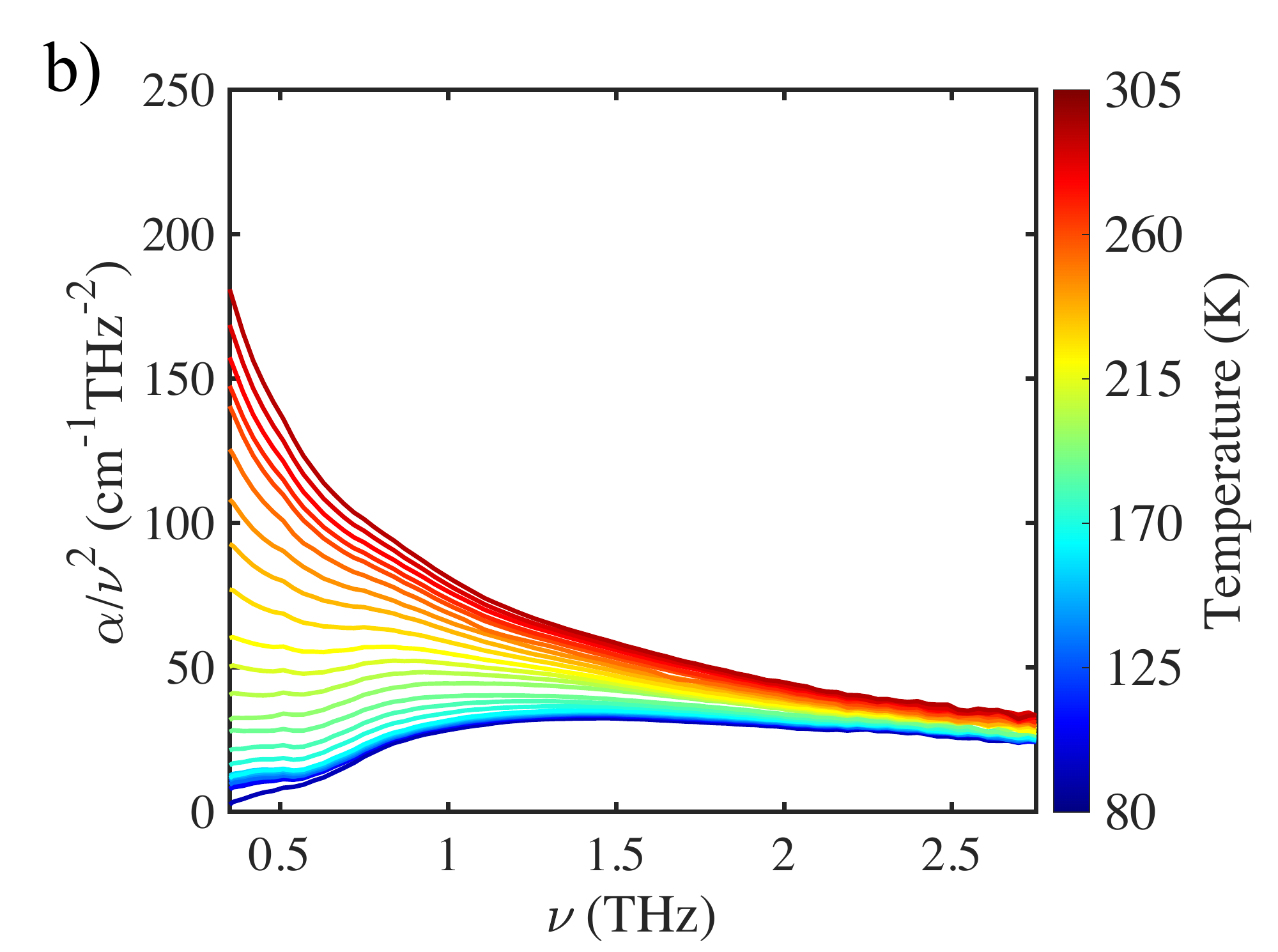}
	\includegraphics[width=0.9\columnwidth]{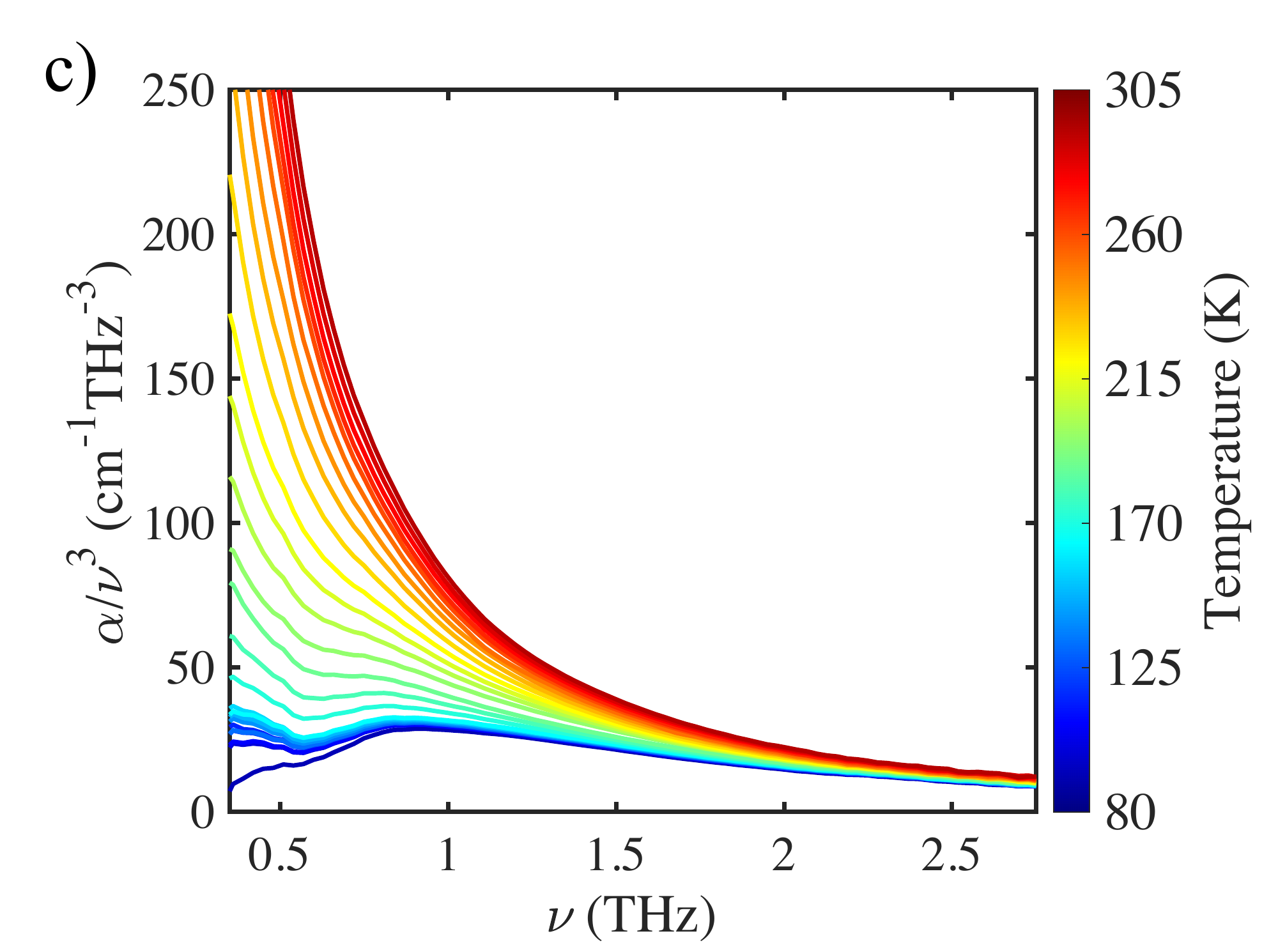}
    \includegraphics[width=0.9\columnwidth]{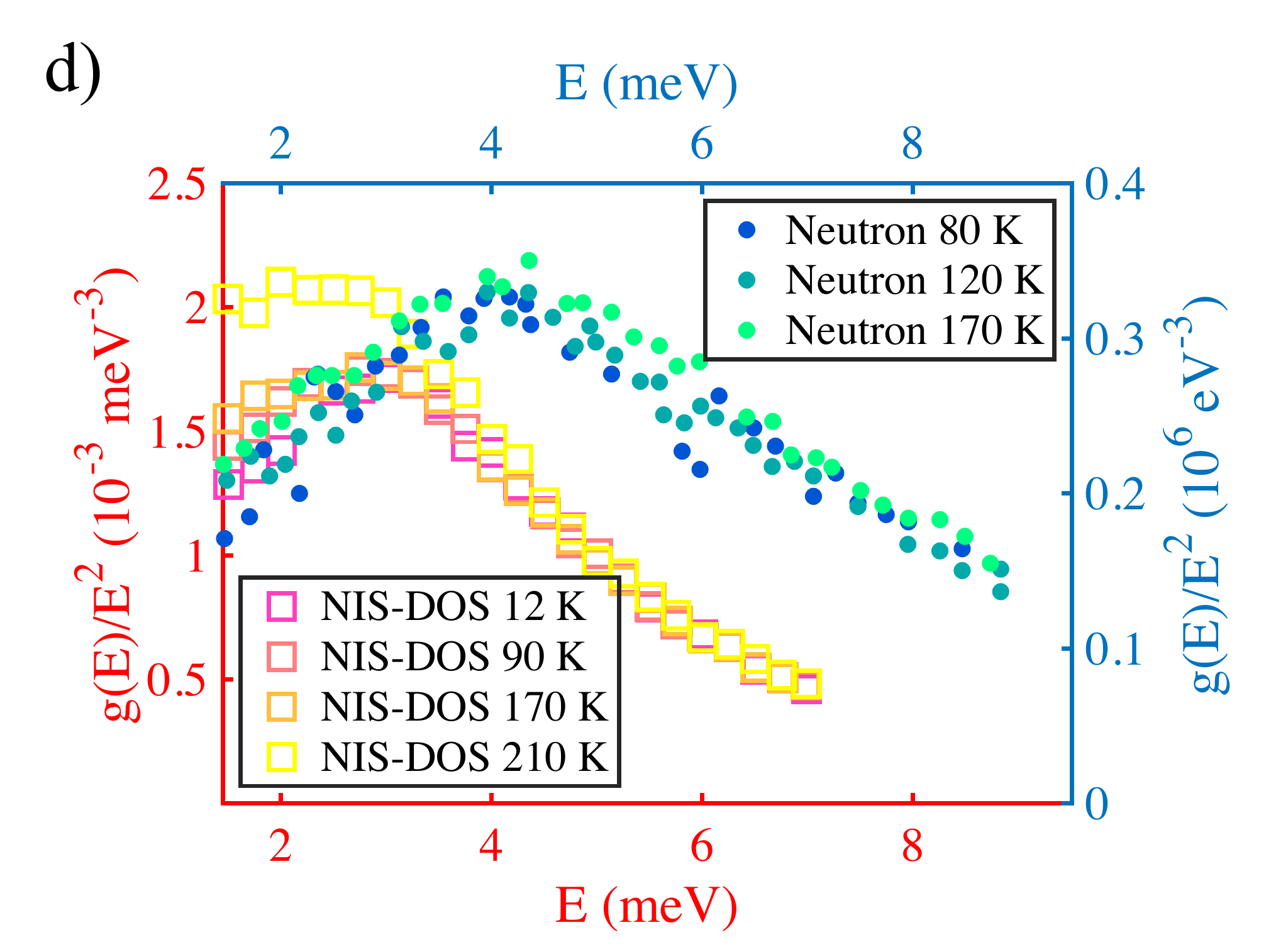}
  	\includegraphics[width=0.9\columnwidth]{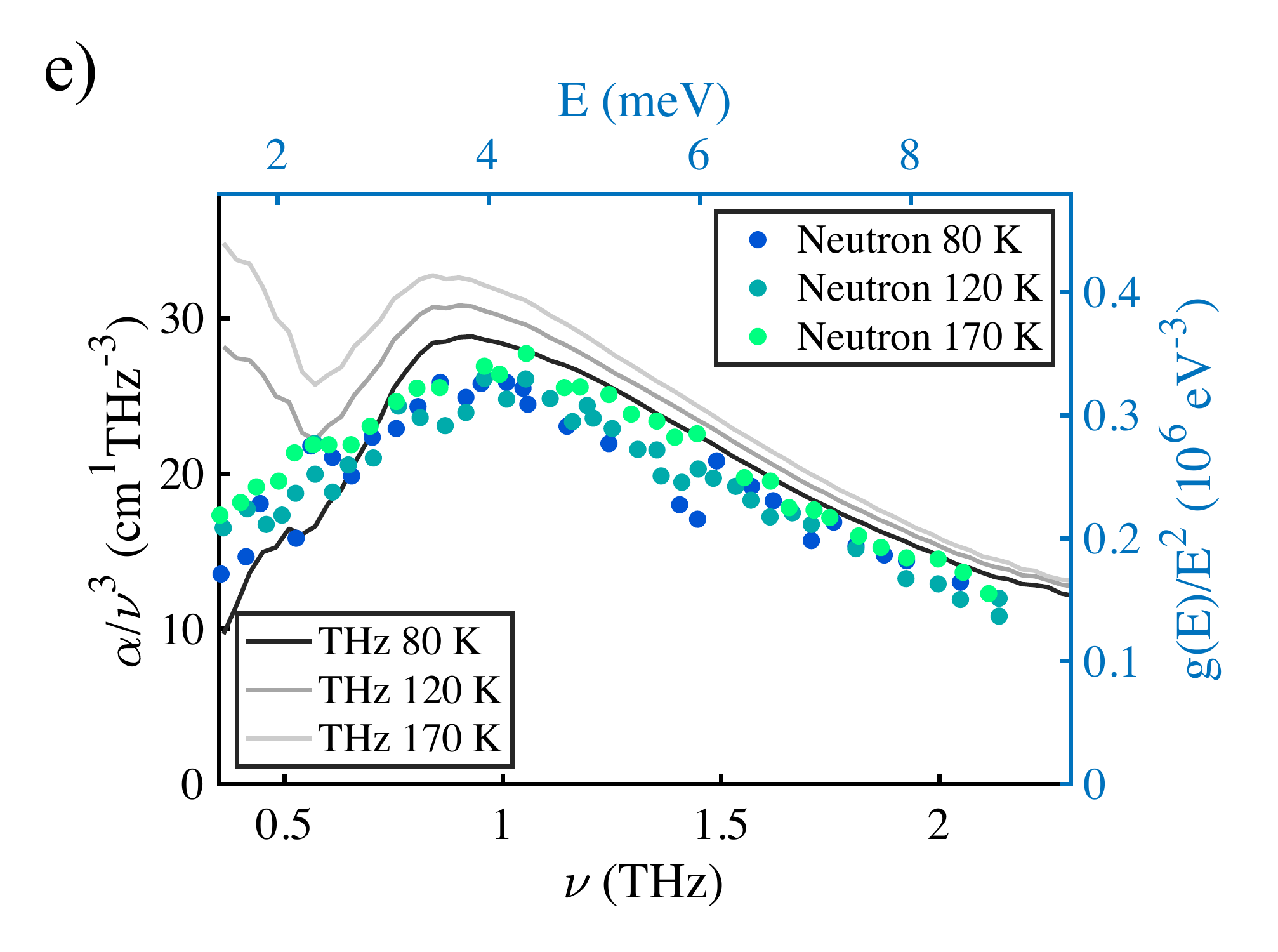}
   	\includegraphics[width=0.9\columnwidth]{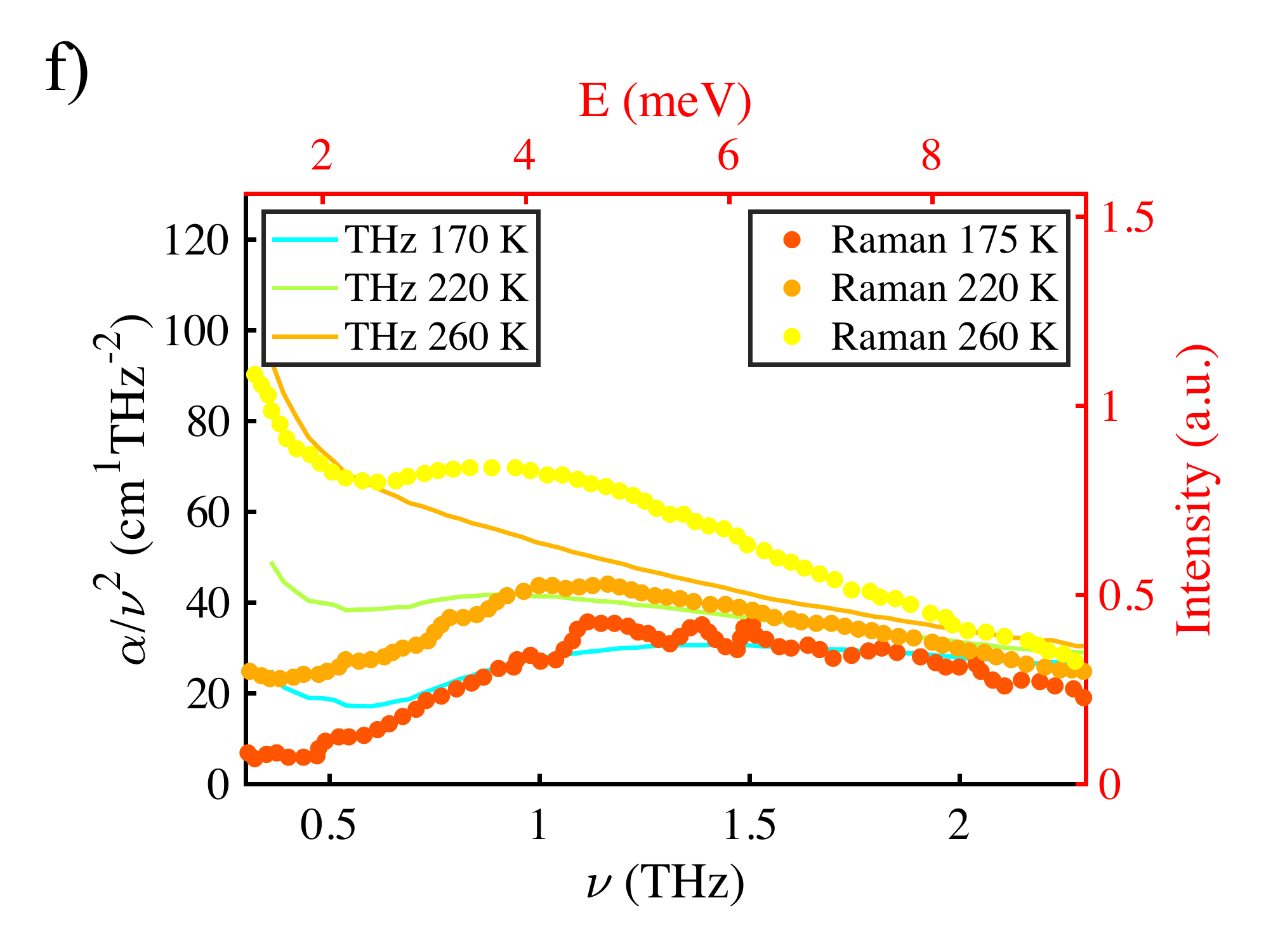}
   	\caption{Comparison of terahertz to neutron, nuclear inelastic scattering,  and Raman data. 
   	(a) THz-TDS absorption spectrum of glycerol for temperature steps of \SI{10}{K} from \SIrange{80}{280}{K} and steps of \SI{5}{K} above. The sample is amorphous over the range of temperatures measured and absorption increases with temperature at all frequencies. 
  	(b) and (c) BP visualised of the same data by plotting $\alpha/\nu^2$ (b) and $\alpha/\nu^3$ (c). A broad maximum is visible at around \SI{1.3}{THz} at low temperatures that disappears upon heating. 
   	(d) Comparison of the BP in the VDOS measured with neutron scattering (\cite{wuttke1996structural}, dots) and the BP measured with nuclear inelastic scattering  (\cite{chumakov2004collective}, squares). The latter maximum is located at slightly lower frequencies than in the neutron scattering data. 
   	(e) Comparison of the BP in the VDOS measured with neutron scattering (\cite{wuttke1996structural}, dots) and $\alpha/\nu^3$ measured with THz-TDS (lines). The maxima and shapes are very similar. 
    (f) Comparison of the $\alpha/\nu^2$ measured with THz-TDS (lines) and the VDOS measured with Raman scattering (\cite{uchino1996low}, dots).  The maxima below \SI{260}{K} occur at similar frequencies and the shape is very similar.
  	\label{fig:f1}}
\end{figure*}

The terahertz absorption spectra measured by THz-TDS of the investigated glycerol-water samples are featureless. An example spectrum is shown in Figure~\ref{fig:f1}a. Without further analysis it is not possible to distinguish between the glass and the liquid states. 

For comparison of the infrared data with neutron and Raman scattering data, the absorption coefficient divided by frequency to the power of two and three is plotted in Figures~\ref{fig:f1}b and c, respectively. At low temperatures, the BP is clearly visible at around \SI{1}{THz} and its amplitude and centre frequency are mostly constant. With increasing temperature, the maximum appears to shift to lower frequencies before features of the BP can no longer be detected at even higher temperatures. 

Chumakov \textit{et al.} \cite{chumakov2004collective} used  nuclear inelastic scattering to measure delocalised collective motions in glycerol with a correlation length greater than \SI{20}{\angstrom}. In comparison to the measurements by Wuttke \textit{et al.} \cite{wuttke1994neutron} the nuclear inelastic scattering data detects the BP at slightly lower frequency, while exhibiting a similar shape overall (Figure~\ref{fig:f1}d). In pure glycerol Wuttke et al. found a pronounced BP at \SI{170}{K} that disappeared at \SI{210}{K}. This observation is confirmed by our THz-TDS measurements (Figure~\ref{fig:f1}c).  Even at very low temperatures (below \SI{120}{K}), the BP, as measured with THz-TDS, increases in intensity while the maximum frequency stays constant. In the same temperature interval, the neutron data shows no temperature-dependent change.

The BP was also visualised with low-frequency Raman scattering by Uchino and Yoko \cite{uchino1996low}. The temperature range in their study was somewhat higher but the peak was found situated at frequencies very close to the ones measured by THz-TDS as shown in Figure~\ref{fig:f1}f. At \SI{170}{K}, the intensity at frequencies below the frequency of the maximum is very similar. It is notable that while in the Raman data a distinct peak with a well-defined maximum was observed at \SI{260}{K}, no distinct peak shape is obvious in the terahertz data.

\begin{figure}
\centering
     \includegraphics[width=0.9\columnwidth]{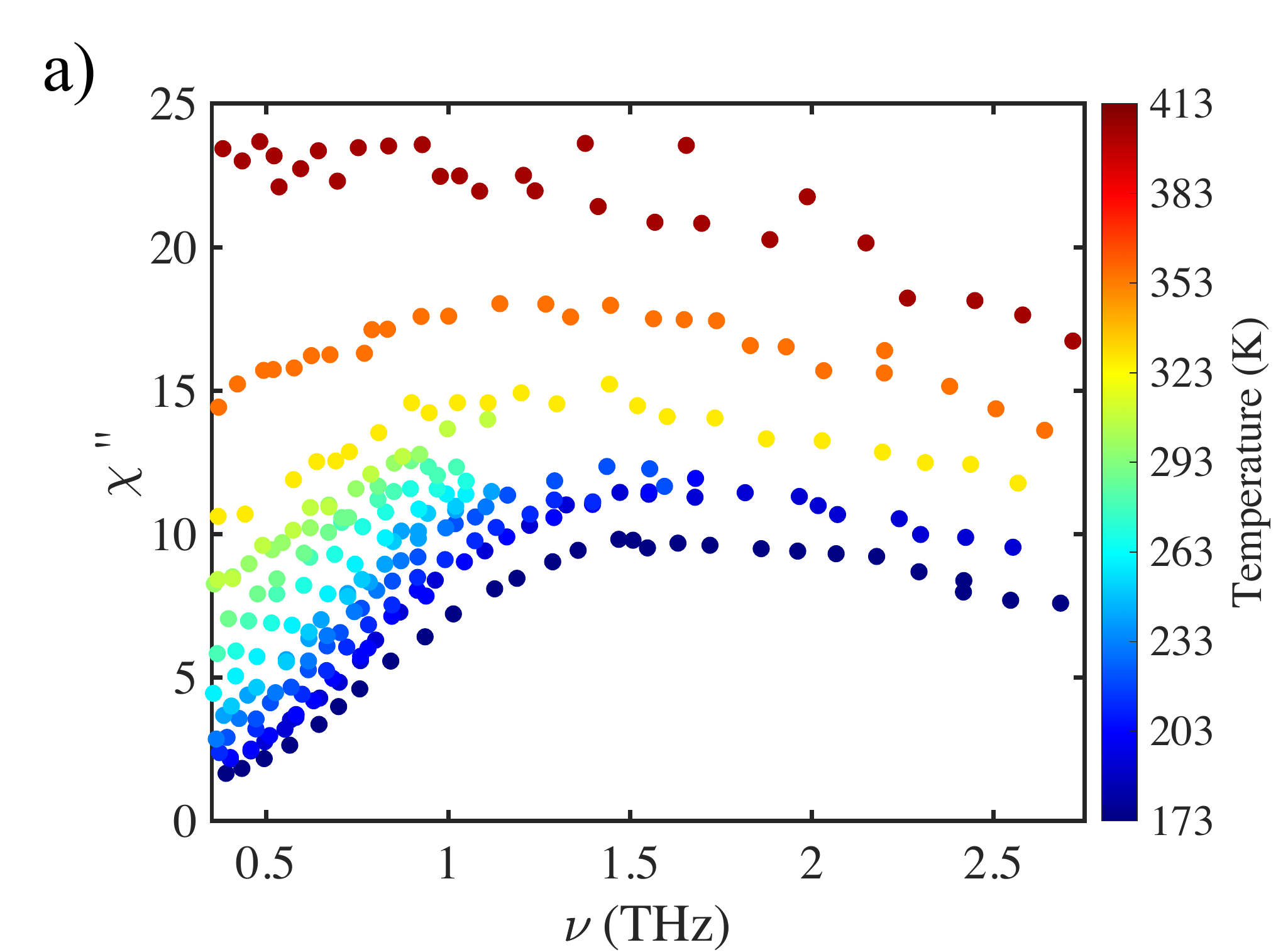}
     \includegraphics[width=0.9\columnwidth]{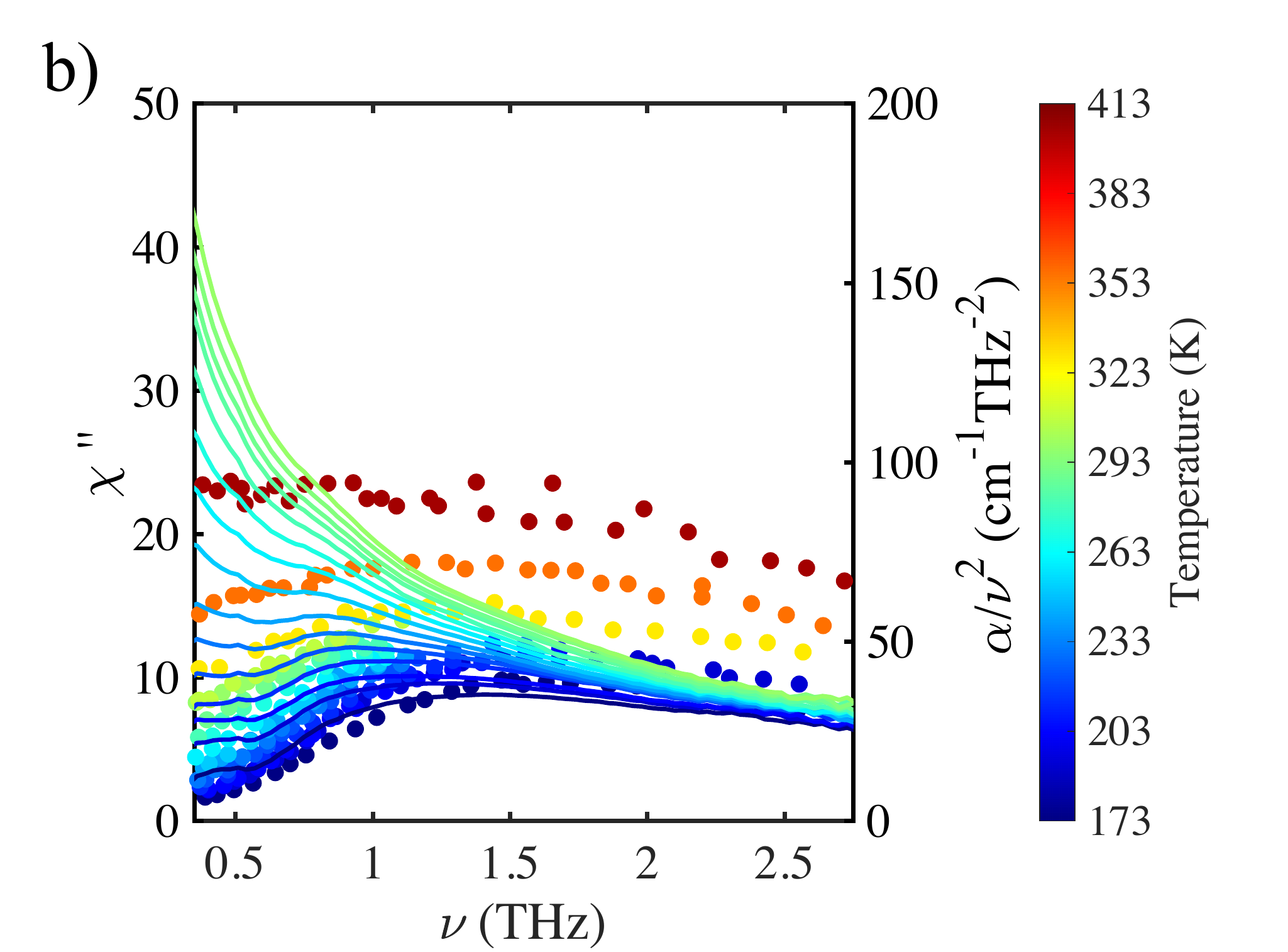}
     	\caption{(a) $\chi''$ from Cummins's light scattering data \cite{wuttke1994neutron} for various temperatures. (b) Comparison with terahertz data. Better agreement with $\alpha/\nu^2$. The maximum in the light scattering data is located at slightly higher frequencies than in the terahertz data.
  	\label{fig:cummins}}
\end{figure}

Light scattering data acquired by Cummins \textit{et al.} \cite{wuttke1994neutron} also show the BP located at approximately \SI{1.5}{THz} at temperatures above \SI{173}{K}. In their data, the BP does not disappear completely, even at elevated temperatures in the liquid regime. The agreement with terahertz data is good for low temperatures, as shown in Figure~\ref{fig:cummins}.

While the BP itself is temperature-independent \cite{schirmacher1998harmonic, marruzzo2013vibrational, schirmacher2015theory}, additional anharmonic effects that contribute to the absorption give the impression of an apparent shift of the BP with temperature. By tracking the changes of the apparent maximum we can observe when the anharmonic effects influence the absolute maximum intensity. This process ultimately results in the intensity of the BP being subsumed entirely by the anharmonic contributions to the absorption intensity.

\begin{figure*}
\centering
     \includegraphics[width=0.9\textwidth]{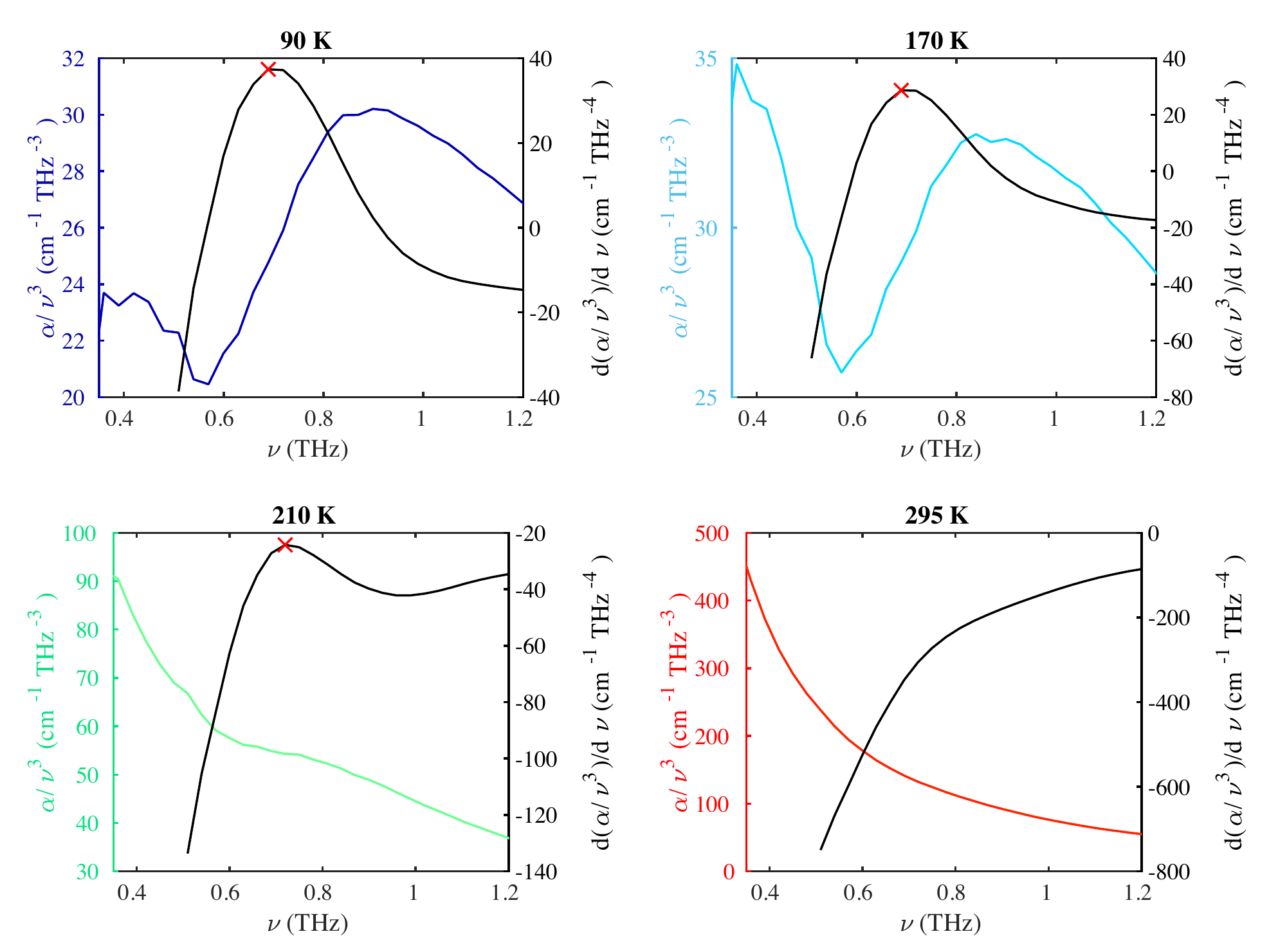}
     	\caption{The density of states and its derivative for different temperatures. The inflection point is highlighted with a cross in the derivative. At \SI{295}{K}, the BP is obscured by anharmonic effects and neither an inflection point nor a maximum in the DOS can be found.
  	\label{fig:f2_new}}
\end{figure*}

\begin{figure}
\centering
     \includegraphics[width=0.9\columnwidth]{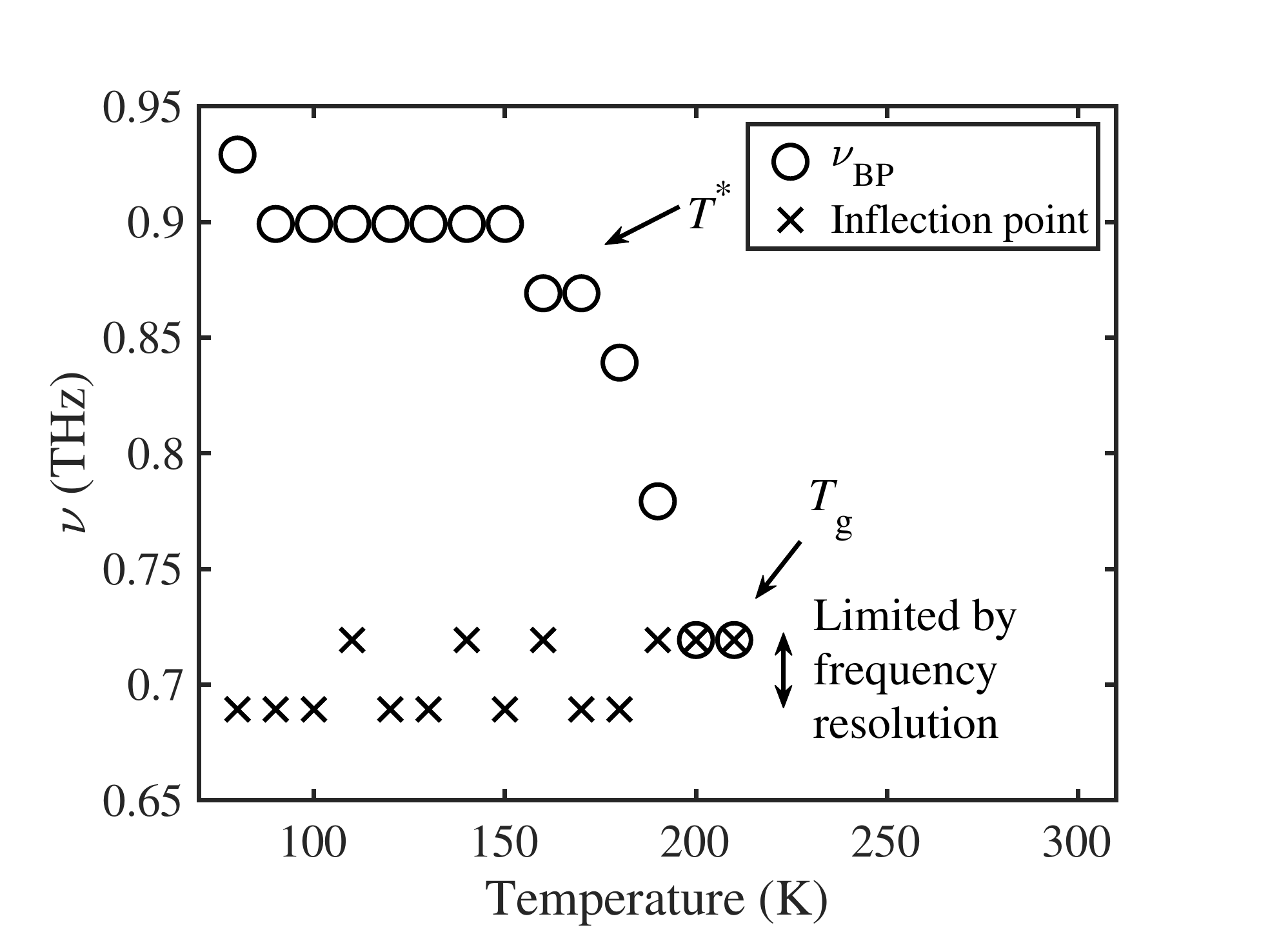}
     	\caption{$\nu_{\text{BP}}$ (circles) and inflection point (crosses) for glycerol for different temperatures as determined  from smoothed data. Above \SI{210}{K}, no inflection points and maxima were found. The frequency resolution (approximately \SI{30}{GHz}) in the smoothed data limits the accuracy when determining $T^*$ and \Tg.
  	\label{fig:f3_new}}
\end{figure}

To track the apparent change in the peak frequency ($\nu_{\text{BP}}$) and its intensity, the data are first smoothed with a moving average and the numerical derivative is calculated. To determine whether the BP is still discernible at elevated temperatures, the inflection point in the DOS is also calculated from the derivative and is shown in Figure~\ref{fig:f2_new} for different temperatures. $\nu_{\text{BP}}$ and the inflection point are plotted against temperature in Figure~\ref{fig:f3_new}.
In pure glycerol, the decrease in the apparent centre frequency of the BP is very sudden. For comparison, the more subtle change in some of the glycerol/water mixtures is shown in Supplementary Figure~S1.  

The behaviour of the BP is hence described by a set of three parameters: the centre frequency at the lowest measured temperature (approximately \SI{0.9}{THz}), the temperature at which the centre frequency is affected by anharmonic effects and appears to decrease ($T^*$, in glycerol approximately \SI{170}{K}), and the highest temperature at which an inflection point is still present in the DOS ($T_g$, in glycerol \SI{210}{K}, Figure~\ref{fig:f3_new}).

\subsubsection{The Absorption Coefficient at Cryogenic Temperatures}

The temperature behaviour of glasses can be separated into 3 regimes: at low temperatures in the harmonic regime, any mobility (e.g. dihedral angle changes) is severely restricted  and molecules vibrate at their equilibrium position due to their thermal energy. 
An increase in thermal energy results in an approximately linear increase in absorption that can be measured throughout the accessible frequency range. This also affects the apparent intensity of the (itself temperature-independent) BP. 
The onset of local mobility at \Tstar\ coincides with a transition region which is concluded by \Tg\ and the anharmonic regime.
With increasing temperature, the VDOS is further enhanced until the liquid state (the third regime) is reached where quasi-elastic scattering occurs and the BP is no longer defined. 

The calorimetric glass transition temperature \Tg\ of glycerol is hard to detect with THz-TDS, as has previously been reported by \citet{sibik2014thermal} and \citet{capaccioli2015coupling}. Sibik \textit{et al.} \cite{sibik2014thermal} found a mean \Tg\ of \SI{194}{K}. 
\citet{capaccioli2015coupling} argued that a secondary glass transition \Tgb\ at \SI{161}{K} was consistent with the same \Tg.

When the spectra can be described by a linear function, that function's gradient $a$  also yields information about the strength of anharmonic processes. 

\begin{figure}
\centering
     \includegraphics[width=0.9\columnwidth]{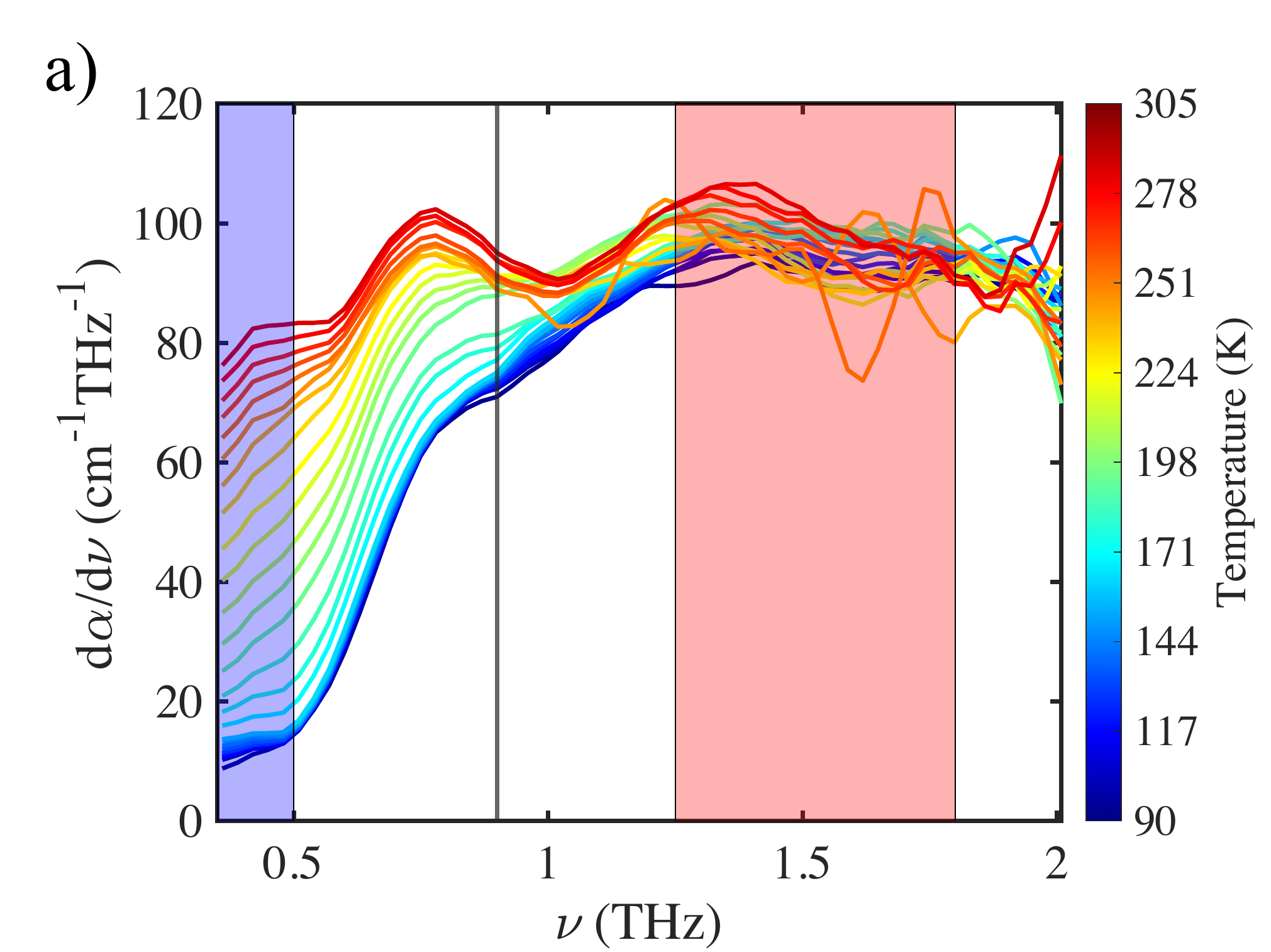}
         \includegraphics[width=0.9\columnwidth]{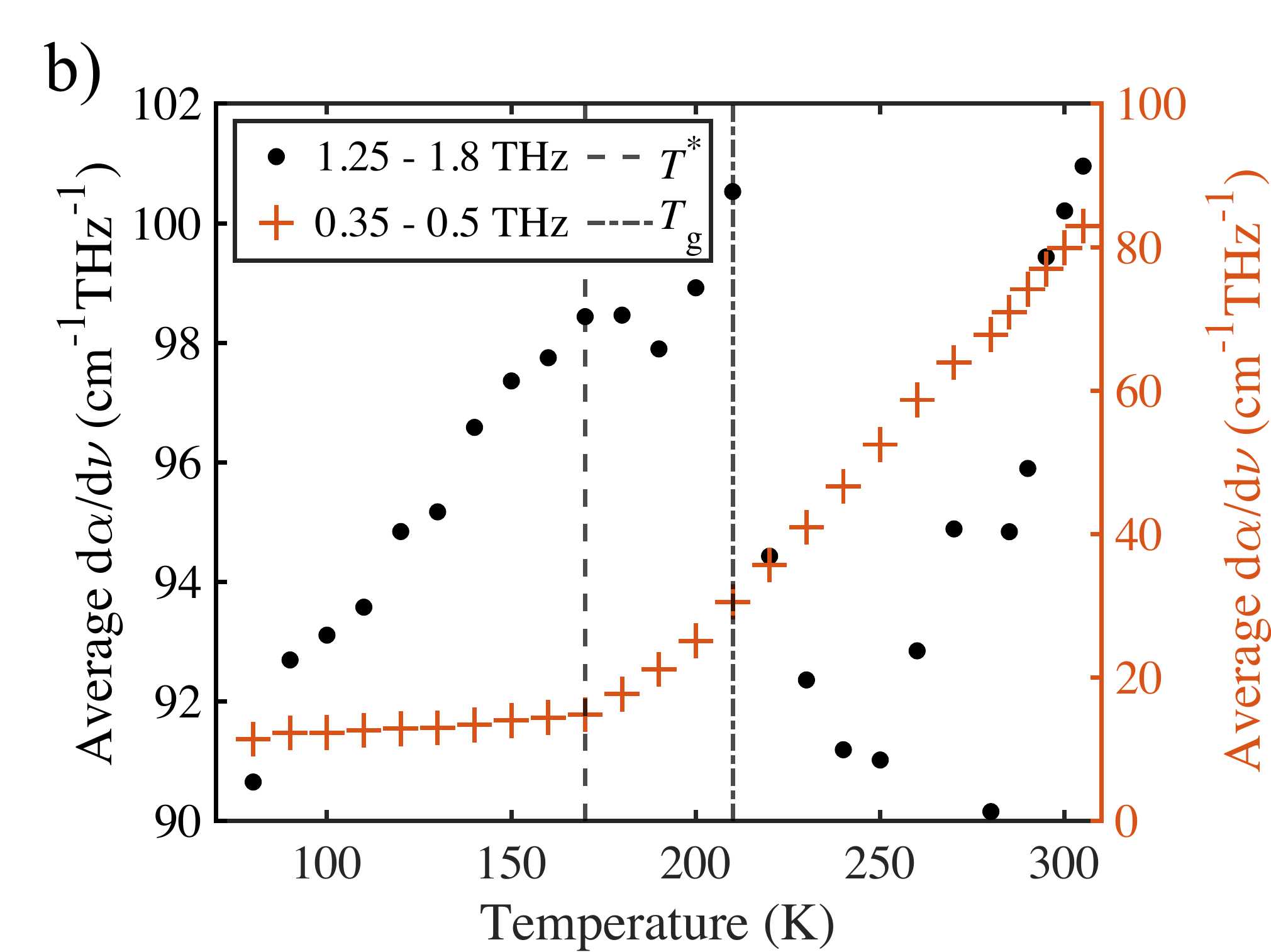}
     	\caption{(a) First derivative of spectra. Above approximately \SI{1.25}{THz} and below \SI{0.5}{THz}, the spectra can be described by linear functions. The vertical line denotes $\nu_{\text{BP}}$. (b) Temperature dependence of average gradient for frequencies between \SIrange{1.25}{1.8}{THz} (dots) and \SIrange{0.35}{0.5}{THz} (crosses), i.e. the anharmonicity parameter $a$. Shown by vertical lines are also $T^*$ and \Tg.
  	\label{fig:dadnu_pure}}
\end{figure}

Whether the absorption coefficient does indeed follow a linear frequency dependence is first investigated by plotting the first derivative, as shown in Figure~\ref{fig:dadnu_pure}a. This shows that the spectra can only be approximated with linear functions below \SI{0.5}{THz} and above \SI{1.25}{THz}. In these regions, the average gradients mirror the behaviour of the $\nu_{\text{BP}}$. Interestingly, lower frequencies are stronger influenced by $T^*$ and higher frequencies by $T_g$.

Below \Tstar, the gradients $a$ at low frequencies are almost constant, implying that anharmonic effects do not dominate. Above \Tstar, however, $a$ increases noticeably. The gradient at higher frequencies is temperature-dependent even below \Tstar. At \Tg, a jump is observed, and $a$ decreases.

These observations and explanations are now used to examine the influence of water content and temperature on the behaviour of glycerol-water mixtures.

\subsection{The Influence of Water Concentration and Temperature on Glycerol-Water Mixtures}

When data at various concentrations are reported, one has to keep in mind that Starciuc \textit{et al.} showed that the transition from predominantly unclustered water molecules to clusters occurs at around 6.1 $\pm$ \SI{0.7}{wt.\percent \ water}. A second threshold was found at 18.6 $\pm$ \SI{4.4}{wt.\percent \ water} and linked to larger, percolating water clusters \cite{starciuc2021water}. At concentrations below the first threshold, water molecules are predominantly unclustered, do not induce major structural changes, and form mostly water-glycerol hydrogen bonds. The LLT found by Murata and Tanaka occurs at higher water concentrations than investigated here \cite{murata2012liquid}. Any change in behaviour found at approximately 6 or \SI{19}{wt.\percent \ water} can therefore be attributed to the influence of water on the structural dynamics by increasing the cluster size and connectivity, as well as a strengthening of the hydrogen bonds between water molecules.

\subsubsection{Harmonic Regime ($T\leq T_g $)}

\begin{figure}
\centering      
\includegraphics[width=0.9\columnwidth]{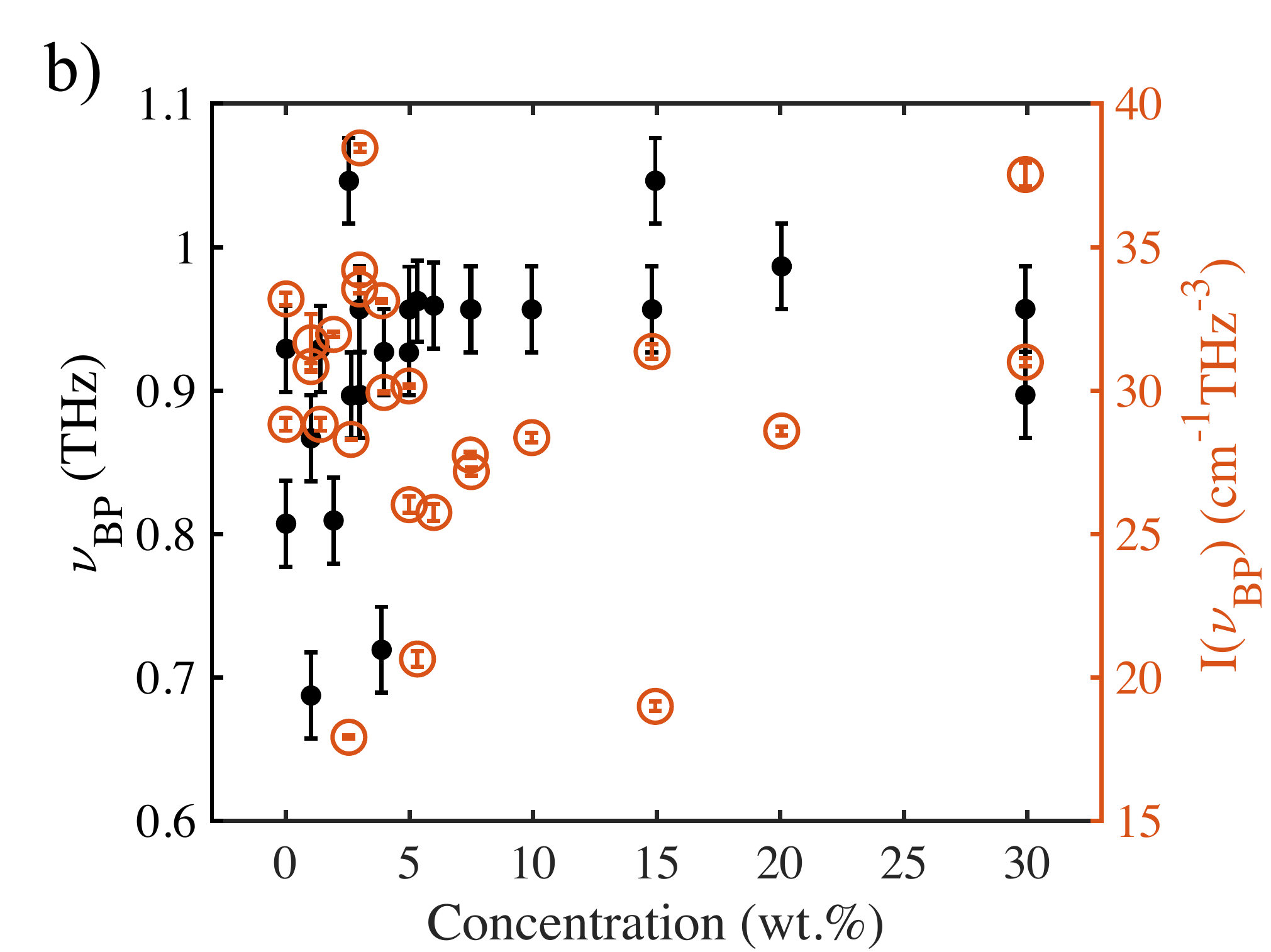}
      \includegraphics[width=0.9\columnwidth]{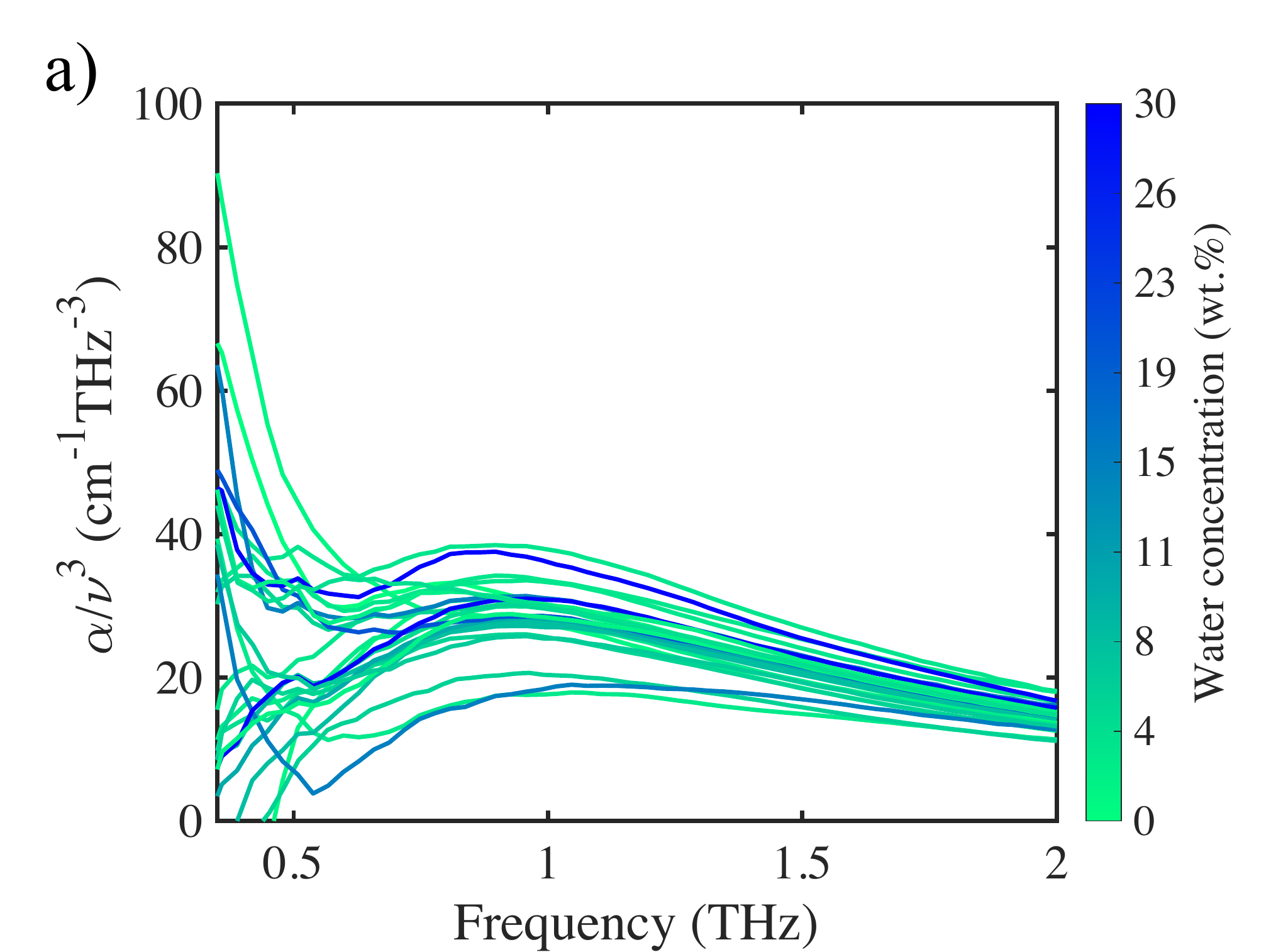}
 	\caption{(a) BP at \SI{90}{K} for all samples. (b) Centre frequency (black dots) and intensity (orange circles) at \SI{90}{K} for all measured concentrations.\label{fig:f6}} 
\end{figure}

The BP is shown for all samples in Figure~\ref{fig:f6}a at a temperature of \SI{90}{K}. The dependence of its centre frequency and intensity on water content are shown in Figure~\ref{fig:f6}b. In most measurements, the BP is centred at \SI{0.95}{THz}. In measurements where the BP is less pronounced, the centre frequency deviates from that value. This seems to be more likely at water concentrations below \SI{5}{\w}.

In the harmonic regime, unclustered and clustered water molecules do not influence the centre frequency of the BP. The Ioffe-Regel crossover is linked to the centre frequency of the BP and denotes a crossover from wave-like to random-matrix-like physics as the mean free path of transverse waves becomes equal to their wavelength  \cite{chumakov2004collective, marruzzo2013vibrational, schirmacher2015theory}. The results indicate that the Ioffe-Regel crossover is not influenced by the presence of clustered or unclustered water.

The onset of water clustering may decrease the intensity of the BP slightly, indicating a higher level of disorder.  While the centre frequency is largely independent of water content, the peak intensity may exhibit a shallow minimum around a concentration of about \SI{5}{\w} when water molecules are homogeneously distributed throughout the sample.

\begin{figure}
\centering
  \includegraphics[width=0.9\columnwidth]{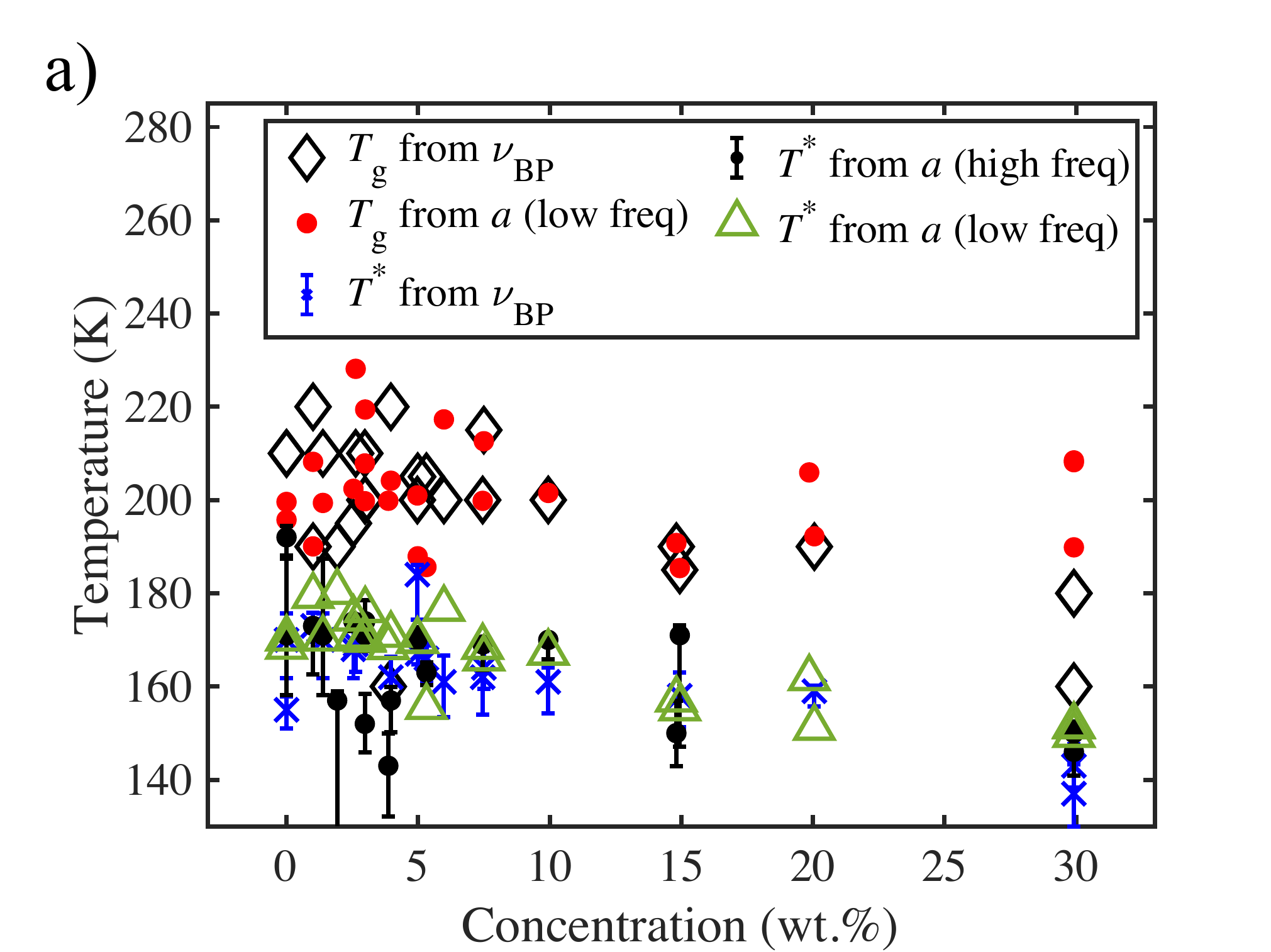}
    \includegraphics[width=0.9\columnwidth]{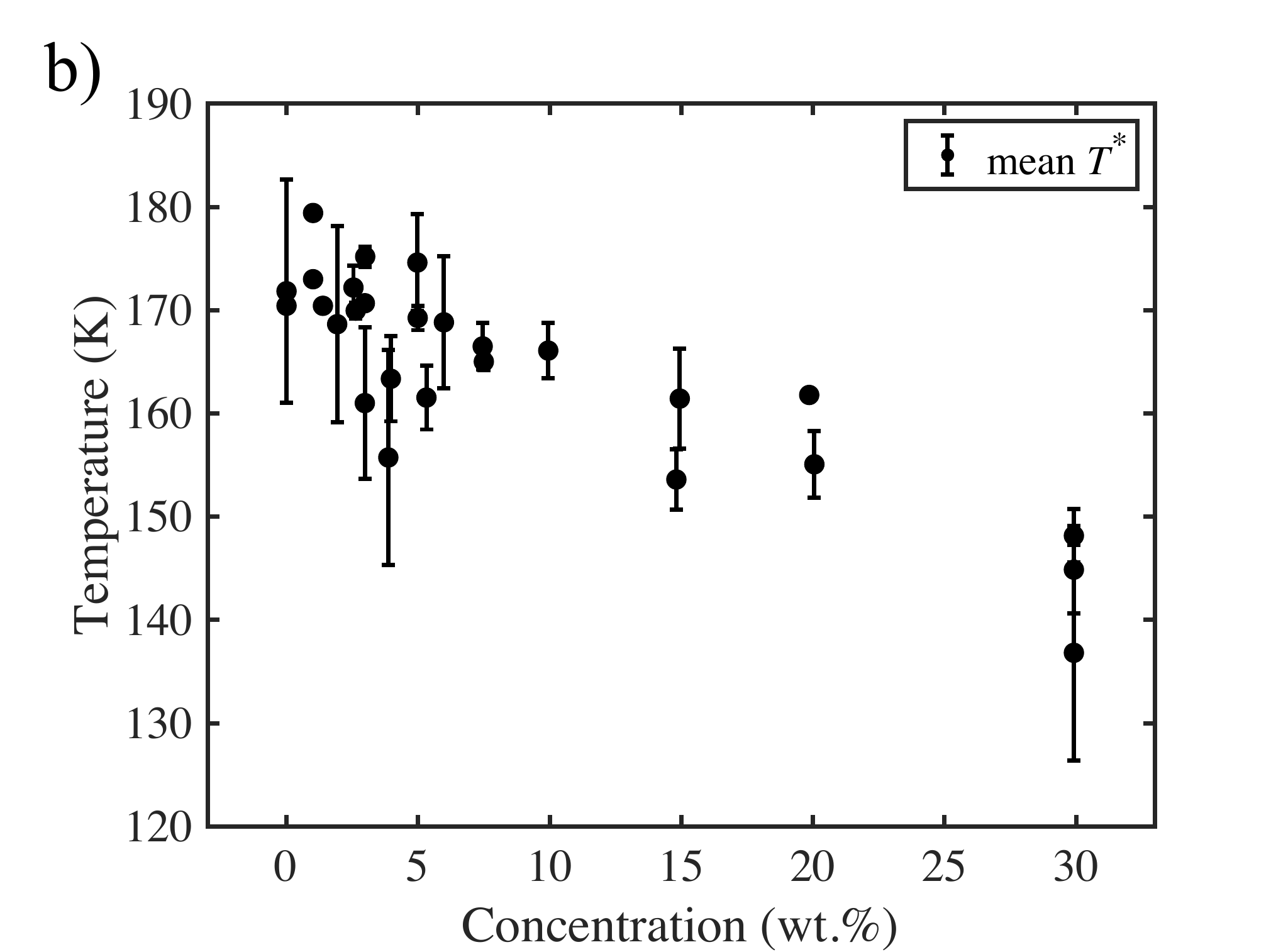}
        \includegraphics[width=0.9\columnwidth]{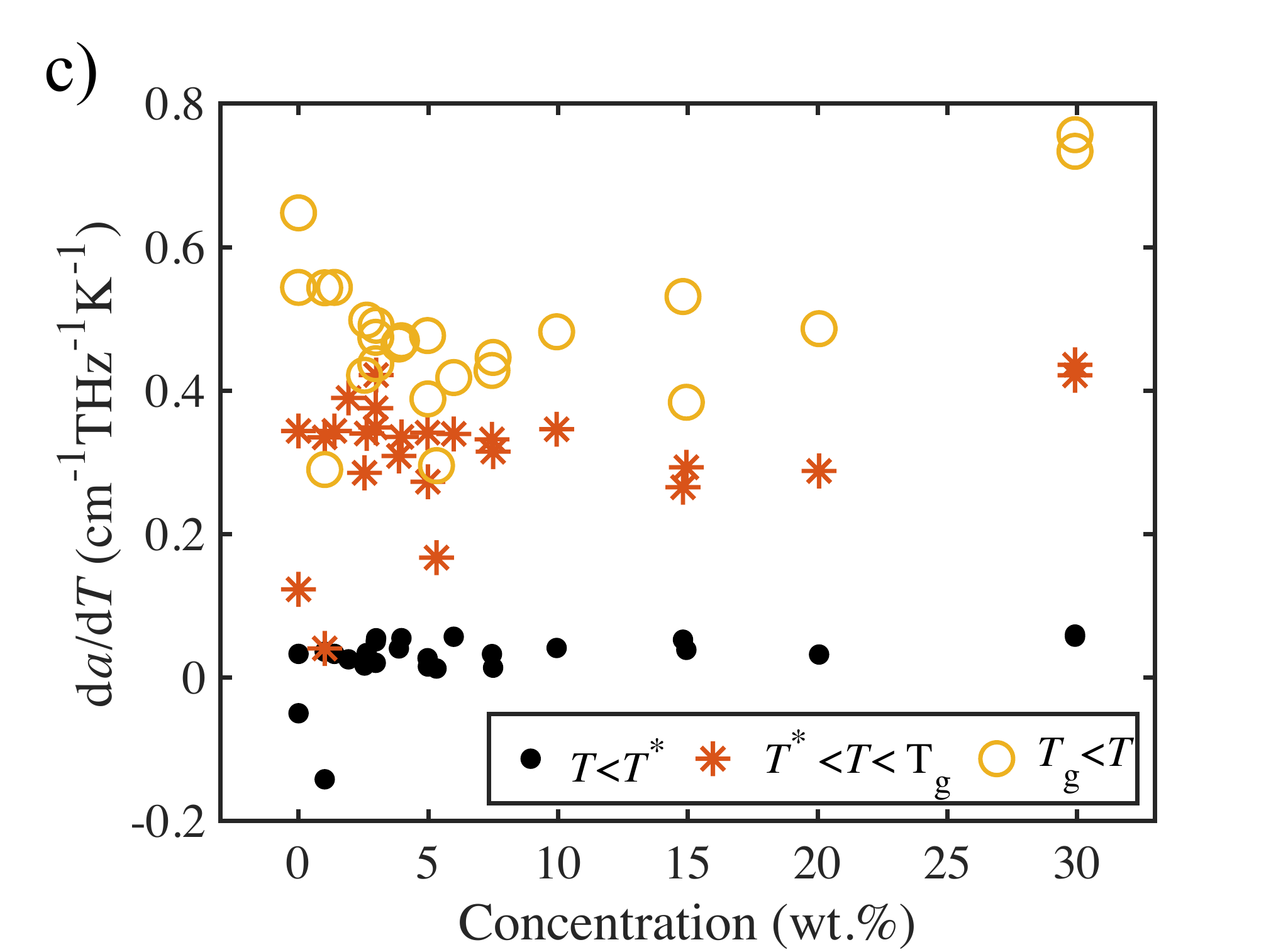}
	\caption{(a) Different transition temperatures characterising the onset of mobility and anharmonicity and the temperature at which the BP dissolves. 
	(b) Mean \Tstar\ for different water contents.
	(c) Rate of change of the anharmonicity parameter with temperature in different temperature regimes. Above \Tg, a shallow minimum is observed around \SI{5}{\w}. \label{fig:f7}} 
\end{figure}

In Figure~\ref{fig:f7}a, an overview of the different transition temperatures is given for all glycerol-water mixtures measured for transition temperatures based on the different methods introduced when discussing pure glycerol. 
The temperatures at which the BP appears to shift and dissolves are characterised for all samples and the results are also shown in Figure~\ref{fig:f7} together with the transition temperatures that are extracted from the anharmonicity parameter $a$ in low and high frequency bands. 

For clarity, the different \Tstar\ have been combined into a mean \Tstar, shown in Figure~\ref{fig:f7}b, and it is apparent that it does not depend on water content below approximately \SI{5}{\w} and decreases at higher concentrations.

This means that unclustered water molecules do not change the temperature at which anharmonic effects result in an apparent shift of the BP. The height of the energy barrier separating the harmonic from the anharmonic regime is hence not influenced by unclustered water molecules embedded in the glycerol matrix.

The shape of the deepest minimum of the PES is independent of water concentration or the presence of water clusters, as otherwise the rate of  change of the anharmonicity parameter $a$ would be different, as shown in Figure~\ref{fig:f7}c. A change in $a$ with temperature corresponds to increased anharmonic contributions.

At water concentrations above \SI{5}{\w}, parameters change their behaviour: the intensity of the BP increases slightly with concentration, and anharmonic effects decrease the temperature at which they influence the appearance of the BP. 

A neutron scattering study by Towey \textit{et al.} \cite{towey2011structure,towey2012molecular, towey2012structural} has shown that at a concentration of \SI{5}{\w}, water monomers are distributed homogenously throughout the material. As the water concentration is increased, water clusters coexist with unclustered water molecules. Past studies suggest that the onset of mobility in amorphous water may lie in the range of \SIrange{120}{150}{K}\cite{ewing2015low} and the PES of bulk-like water is hence expected to comprise shallower minima than that of glycerol where the onset of mobility lies at higher temperatures. The THz-TDS results indicate a shift in the structural dynamics once clusters form, accompanied by a reduced height of the potential energy barrier separating the harmonic from the anharmonic regime.

As the water concentration is increased up to \SI{16.4}{\w} water, the percentage of water molecules that are part of clusters increases to \SI{80}{\percent} \cite{towey2011structure,towey2012molecular, towey2012structural}. In our THz-TDS measurements we do not see a difference between \Tstar\ for mixtures containing \SI{15}{\w} and \SI{20}{\w}. This indicates that the presence of water-water hydrogen bonds lowers the barriers on the PES, but once clusters dominate, the intensity of anharmonic effects does not depend on the number of clusters, but on their mean size.  

Water-water cooperative domains \cite{hayashi2006slow, sudo2002broadband} and percolating water clusters \cite{starciuc2021water} have been found at even higher water content. In our THz-TDS measurements, the onset temperature of anharmonic effects is decreased noticeably in samples with a water concentration of \SI{30}{\w}. We can hence infer that the mean size of clusters influences the PES and thereby structural dynamics stronger than their number and that the height of the lowest energy barrier is the limiting factor determining the onset temperature of anharmonicity. Regions with lower energy barriers first exhibit an increase in mobility which is measured by THz-TDS.

\subsubsection{Anharmonic Regime ($T_g<T<T_m$)}

Below \Tstar, the rate of absorption change does not strongly depend on water content,  as shown in Figure~\ref{fig:f7}c. 
Above \Tg, the rate of absorption change starts to increase with water concentration above \SI{5}{wt.\percent \ water} (see Figure~\ref{fig:f7}c). There is also evidence for a shallow minimum at around \SI{5}{\w} which could be due to a very homogenous sample including many unclustered water molecules.

Unclustered water reduces the rate of change of the anharmonicity parameter at a water content below \SI{30}{\w} compared to pure glycerol. This ``plasticiser effect'' could be because the unclustered water molecules are surrounded by larger glycerol molecules, and any reorientation of the glycerol molecules first requires the breaking of hydrogen bonds with water.
Once clusters are formed, however, the larger mobility of water molecules in a bulk-like environment increases the rate again.

The higher the temperature and mobility, the more anharmonic effects obscure the BP until it can no longer be resolved. \Tg\ does not depend on water content below \SI{10}{\w}, and decreases for water concentrations up to \SI{30}{\w}. This indicates that the presence of a number of larger water clusters increases anharmonic effects. Unclustered water and small isolated clusters have no effect.

Raman scattering experiments have shown that an underlying BP continues to be present above \Tg\  \cite{uchino1996low} which can no longer be resolved by THz-TDS because it is obscured by anharmonic effects. 

The calorimetric \Tg\ as measured with DSC shows a dependence on water concentration even below \SI{10}{\w}\cite{starciuc2021water}. This can be attributed to the different criteria for determining the exact value. While the calorimetric \Tg\ is related to a change in heat flow, our definition based on anharmonic effects is therefore expected to result in slightly different values.

At a temperature of approximately \SI{30}{K} above \Tg\ lies the melting temperature of the crystalline system and by further heating the supercooled liquid, it enters the liquid regime.

\subsubsection{Liquid Regime ($T=\SI{293}{K}$)}

A room temperature (\SI{293}{\kelvin}) set-up was utilised to measure glycerol-water mixtures in the same concentration range (\SIrange{0}{30}{wt.\percent \ water}) as discussed above. With this setup it was easy to repeat measurements very rapidly and the spectra shown in Figure~S4a are the average of 18 measurements each. The absolute error associated with the absorption coefficient measurement is hence decreased and the data quality and spectral range are better than for measurements performed at variable temperatures. 

At room temperature, the mixtures are fully liquid and no BP is present. The absorption coefficient at \SI{1}{THz} depends linearly on water concentration $c$, as can be seen in Figure~S4b: 
\begin{equation}
	\alpha(c)=\SI{1.47}{\per\centi\meter\per\percent}\cdot c[\w]+\SI{65.6}{\per\centi\meter}
\end{equation}
and the presence or absence of water clusters does not seem to influence this relationship. To investigate how water clusters influence the THz-TDS spectra of liquid glycerol-water mixtures, the frequency-dependent absorption is hence evaluated. The high data quality allowed to extrapolate the absorption coefficient measured in the range \SIrange{0.6}{2.5}{THz} according to a model developed by Schirmacher \textit{et al.} \cite{chumakov2004collective}. 
However, the model only makes predictions, and in a future experiment, the spectra could be measured on a THz-TDS system with a higher spectral bandwidth.

The better data quality allows to fit a model to the data in the range \SIrange{0.6}{2.5}{THz} to extrapolate the spectra further and investigate the influence of concentration: 
	\begin{equation}
	\alpha=A\cdot\nu^2\cdot\exp{(-\nu/(\nu_{\text{c}}/2))}+C
\end{equation}
Which predicts a broad peak centred at $\nu_{\text{c}}$. 
The model is fitted to experimental data restricted to frequencies below \SI{2.5}{\tera\hertz} because with increasing frequency the noise increases  \cite{jepsen2005dynamic}. Measured spectral data are reduced to the fit parameters and the influence of water concentration is examined.

\begin{figure*}
\centering
  	\includegraphics[width=0.9\columnwidth]{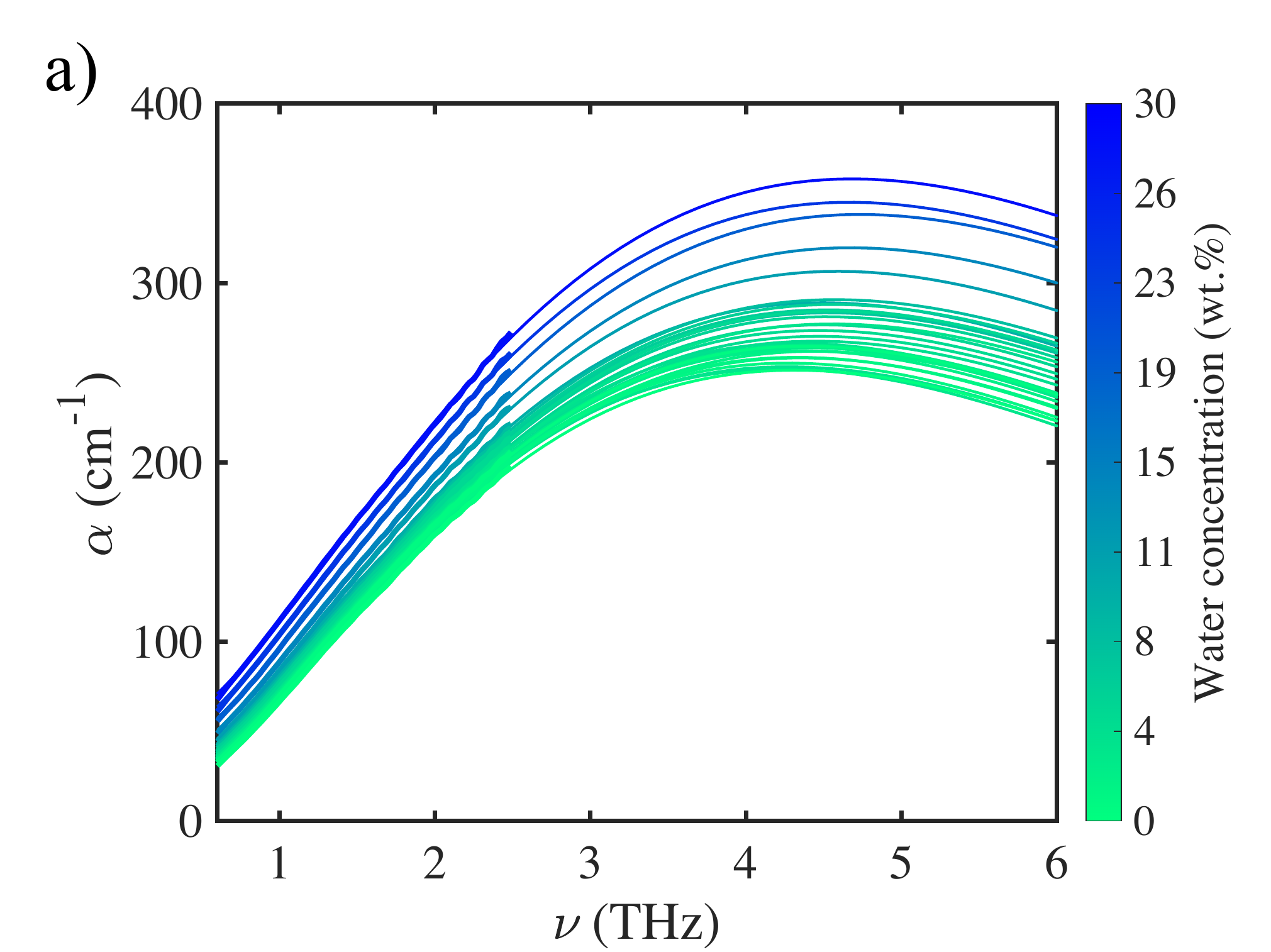}
 	\includegraphics[width=0.9\columnwidth]{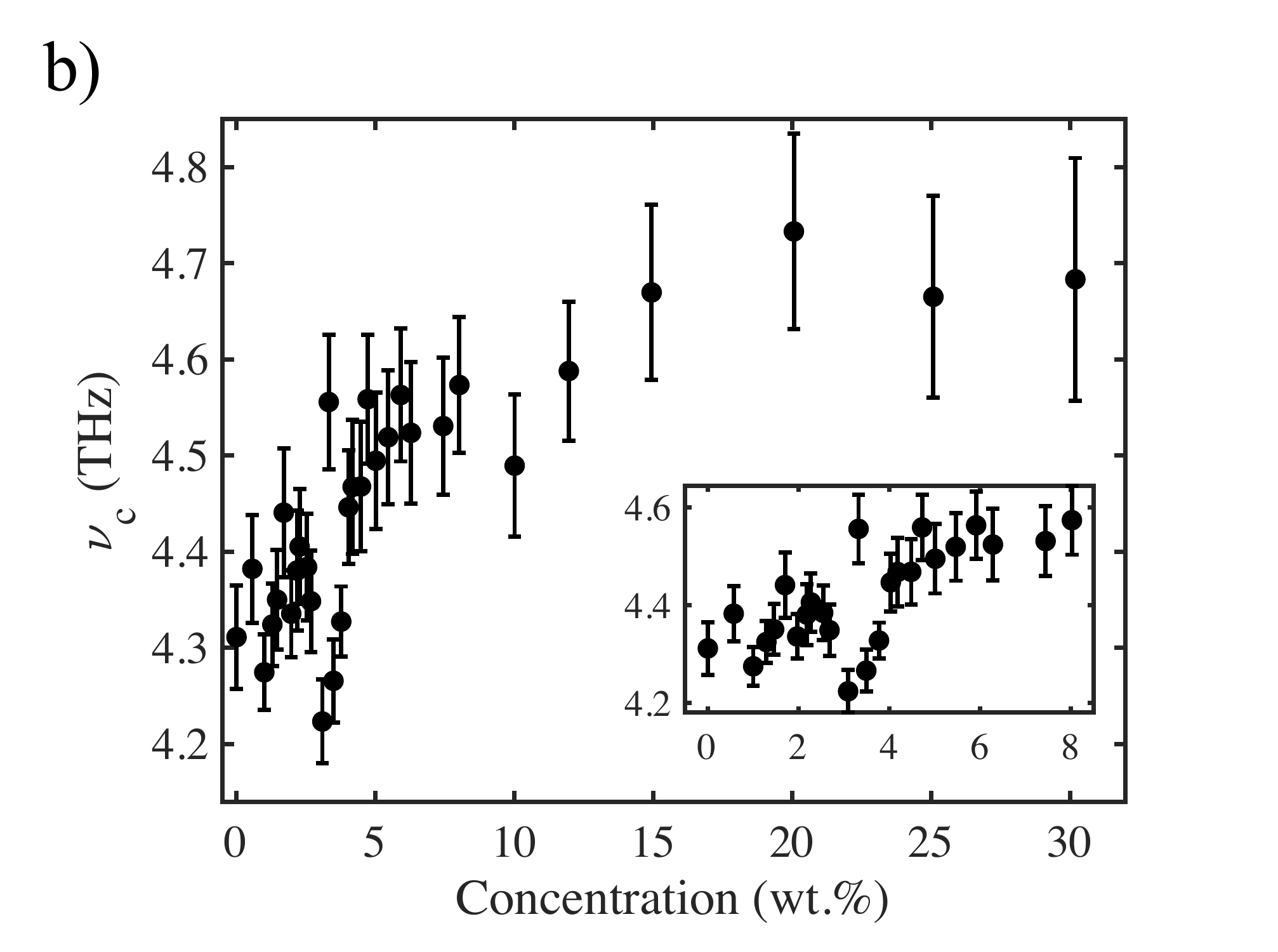}
    \includegraphics[width=0.9\columnwidth]{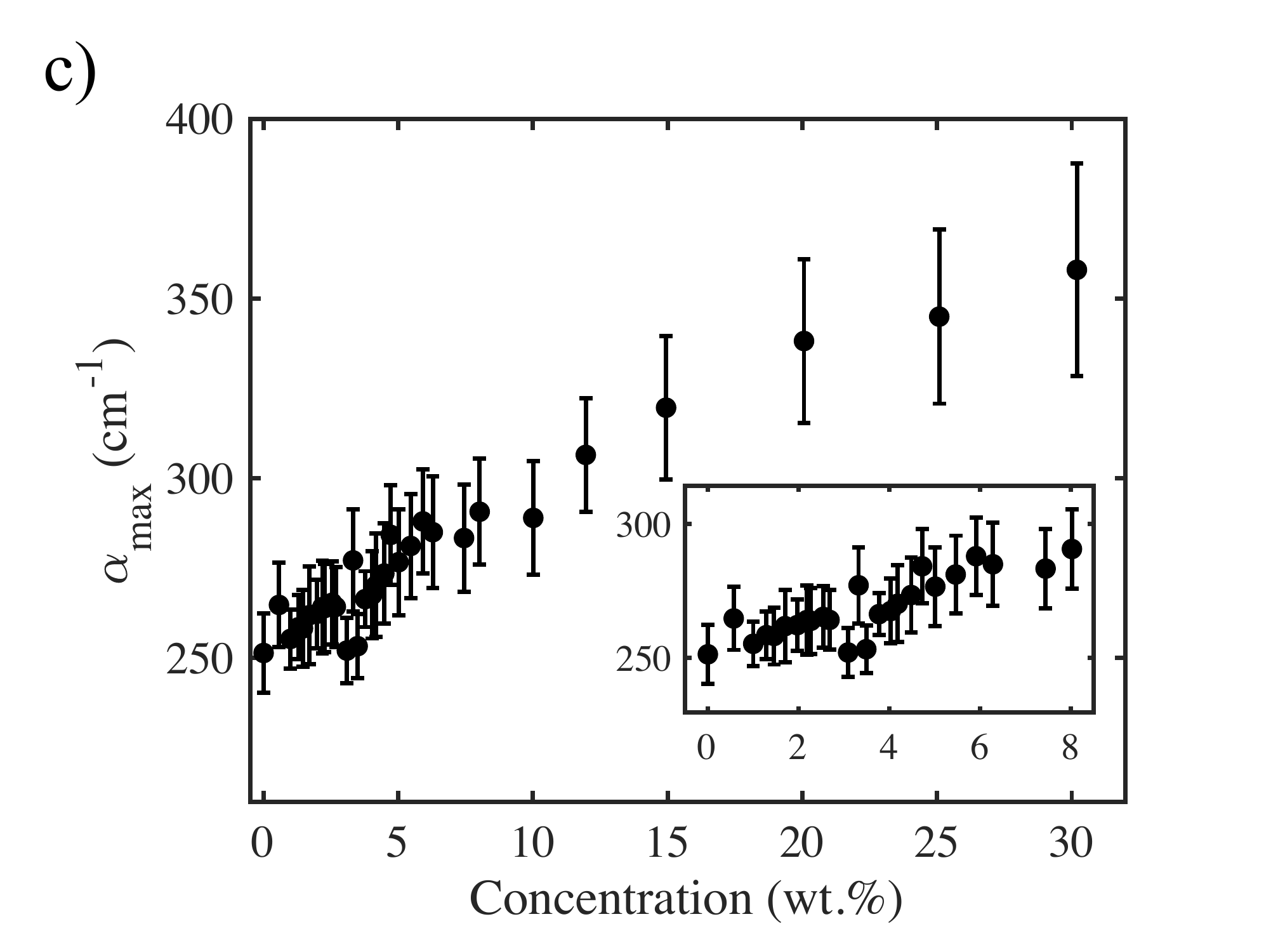}
    \includegraphics[width=0.9\columnwidth]{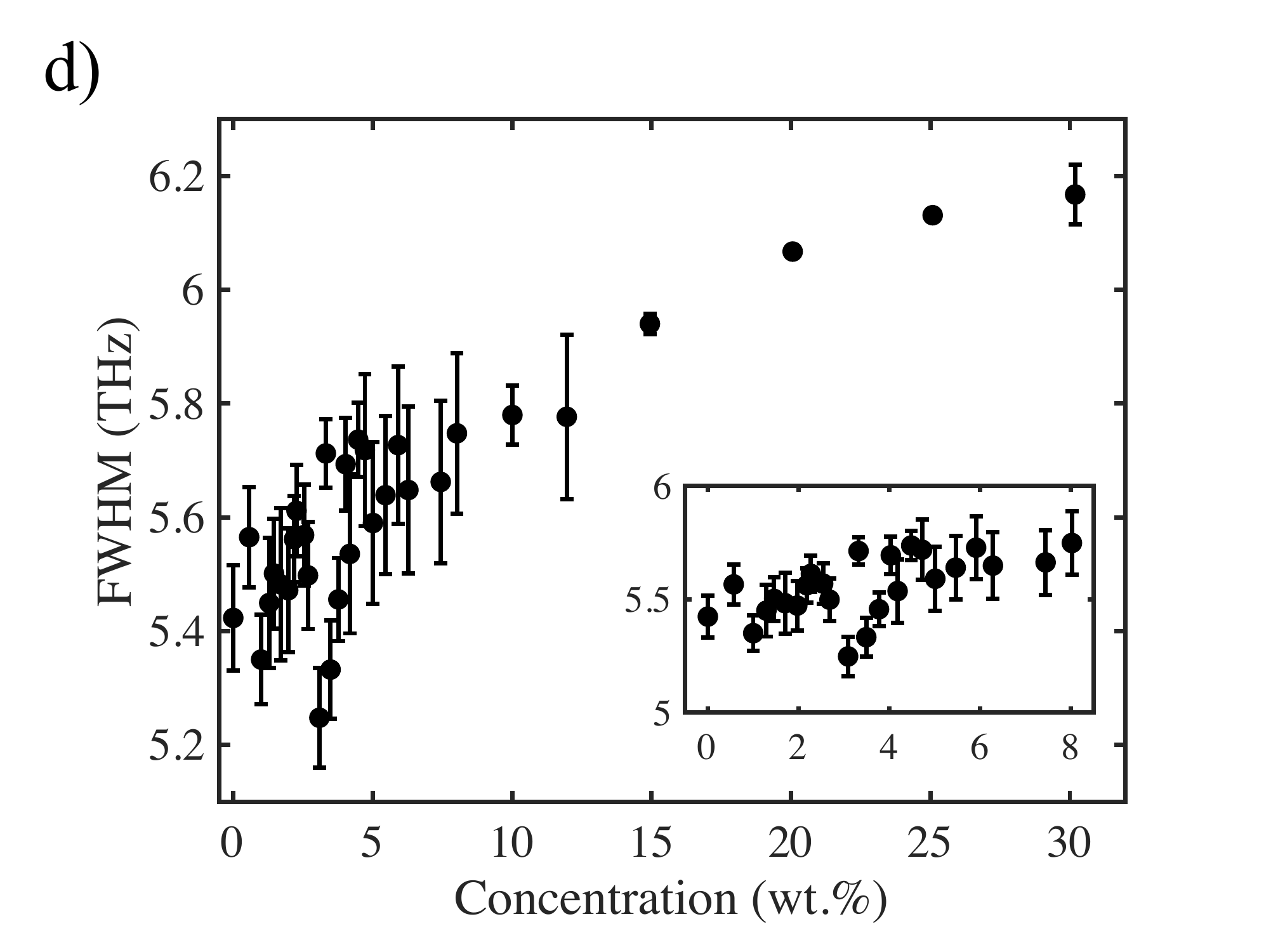}
  	\caption{(a) Measured spectra (up to \SI{2.5}{THz}) and extrapolated spectra at higher frequencies.\newline
  	(b)-(d) Predicted centre frequency (a), maximum absorption (b), and FWHM of the spectral shape for different water contents. Error bars are calculated from 95\% confidence interval of the fit. The insets show the concentration range \SIrange{0}{8.5}{\w} in more detail.	
		\label{fig:RT_VDOS}}
\end{figure*}

The parameters $\alpha_{\text{max}}$ (maximum absorption) and the full width at half maximum (FWHM) describing the spectra are determined for different concentrations as shown in Figure~\ref{fig:RT_VDOS} with the concentration range below \SI{8.5}{\w} water shown in detail in the insets.
As expected, an increase in water content leads to an increase in absorption, as well as to a broadening of the peak.  This also leads to the previously discussed increase in the absorption at \SI{1}{THz}. 

Unclustered water molecules are tightly integrated into the hydrogen-network, with most glycerol hydrogen bonds formed between water and glycerol molecules \cite{starciuc2021water}. Above the glass transition temperature, these water molecules exhibit different degrees of freedom which broadens the spectral feature at higher frequencies.  
Increasing the amount of unclustered water leads to the most noticeable change in those three parameters over the whole concentration range studied. 

At \SI{6}{\w}, the rate of change in maximum absorption with water content decreases slightly. Once water clusters form, the number of water-glycerol hydrogen bonds decreases and the number of water-water bonds increases. Once clusters are present, a higher water content changes the dynamics less because the water would only increase the cluster size and thereby disrupt the hydrogen-bonded network of glycerol molecules less. 
Within the clusters, the water dynamics are markedly different from unclustered water molecules that are hydrogen-bonded to glycerol.

While the second clustering threshold just below \SI{20}{\w} was not clearly observed in the BP, it influences the behaviour of the system at higher frequencies. All three parameters describing the VDOS change less above a water content of \SI{20}{\w}. Most notably, the centre frequency plateaus above \SI{20}{\w}. 
At room temperature, bulk water has a higher absorption at \SI{1}{THz} than glycerol (\SI{220}{\per\centi\meter} compared to \SI{66}{\per\centi\meter}) and the larger the water clusters, the more they behave like bulk water. Once percolating water clusters form, they dominate the absorption of liquid glycerol-water mixtures at higher frequencies comprising the VDOS.

\section{Conclusions}
While the terahertz spectra of amorphous solids and liquids are featureless, they provide considerable information about glass transition temperatures and the related mobility changes. They also offer insight into the vibrational density of states and the BP.

We show that the infrared absorption coefficient measured with THz-TDS can be theoretically related to the reduced Raman intensity ($\propto \alpha/\omega^2$) and the reduced DOS ($\propto \alpha/\omega^3$) and the agreement with experimental results confirms this.

The temperature behaviour is characterised by three regimes: below the glass transition temperature, the system is trapped in a deep minimum of the potential energy landscape and exhibits mostly harmonic vibrations. The onset of local (and, at slightly higher temperatures), global mobility is accompanied by anharmonic excitations, obscuring the BP and eventually leading to its dissolution. Once the system crosses over into the liquid regime, the absorption spectra are dominated by quasi-elastic scattering. 

The implications of this model are  examined in detail on the example of glycerol and the insights are applied to study the influence of clustered and unclustered water.

In the harmonic regime, the shape of the PES is unaltered in the presence of clustered and unclustered water molecules. Unclustered water molecules are embedded into the glycerol matrix and do not decrease the onset temperature of anharmonicity. A change in structural dynamics is observed at a water concentration of approximately \SI{5}{\w}, corresponding to a transition from isolated water molecules distributed homogeneously throughout the sample to the presence of small water clusters and an increased number of water-water hydrogen bonds which lower the barriers on the PES. 

Interestingly, the intensity of anharmonic effects does not depend on the number of water clusters but on their mean size. 

Data acquired at room temperature is extrapolated according to a model by Schirmacher et al. which allows to investigate the spectral shape for different concentrations for liquid glycerol-water mixtures and confirms a change of dynamics once percolating water clusters formed between \SIrange{15}{20}{\w}.

This methodology has great potential to be applied to other systems to investigate the change of mobility and dynamics.

\begin{acknowledgements}
	JK thanks the EPSRC Cambridge Centre for Doctoral Training in Sensor Technologies and Applications (EP/L015889/1) and AstraZeneca for funding. Disclosures: ES is an employee of AbbVie.  He participated in conceptual design of the study and in reviewing and approval of the publication.
\end{acknowledgements}

\end{document}